\documentclass[a4paper, traditabstract, longauth]{aa} 

\usepackage{graphicx}
\usepackage[usenames]{color}

\usepackage[breaklinks, colorlinks, citecolor=blue]{hyperref}
\usepackage{natbib}
\usepackage{longtable,lscape}
\usepackage{txfonts}
\usepackage{amsfonts}
\usepackage{amssymb}
\usepackage{newlfont}
\usepackage{textcomp}
\usepackage{times}
\usepackage{units}
\usepackage{array}
\bibpunct{(}{)}{;}{a}{}{,} 

\graphicspath{./} 

\newcolumntype{C}{@{\,}c@{\,}}
\def\Tab#1{\tabular[t]{@{}c@{}}#1\endtabular}

\def\reff@jnl#1{{\rm#1\/}}
\def\apj{\reff@jnl{ApJ}}       
\def\apjs{\reff@jnl{ApJS}}     
\def\aaps{\reff@jnl{A\&AS}}    
\def\mnras{\reff@jnl{MNRAS}}   
\def\prd{\reff@jnl{Phys.\ Rev.\ D}}    
\def\jcap{\reff@jnl{JCAP}}     

{\begin{enumerate}\setlength{\itemsep}{0mm}}%
{\end{enumerate}}
{\begin{enumerate}\setlength{\itemsep}{0mm}}%
{\end{enumerate}}

\newcommand{\bc}{\begin{center}}
\newcommand{\ec}{\end{center}}
\newcommand{\bi}{\begin{itemize}}
\newcommand{\ei}{\end{itemize}}
\newcommand{\ben}{\begin{enumerate}}
\newcommand{\een}{\end{enumerate}}

\usepackage{mathrsfs}
\usepackage{mathbbol}

\newfont{\gwpfont}{cmssq8 scaled 1000}

\begin{document}

\title{Detecting the integrated Sachs-Wolfe effect with stacked voids}

\author{St\'ephane Ili\'c, Mathieu Langer, Marian Douspis}
\institute{Institut d'Astrophysique Spatiale, UMR8617, Universit\'e Paris Sud \& CNRS, B\^at. 121, Orsay F-91405, France}

\abstract{The stacking of cosmic microwave background (CMB) patches has been recently used to detect the integrated Sachs-Wolfe effect (iSW). When focusing on the locations of superstructures identified in the Sloan Digital Sky Survey (SDSS), Granett et al.\ (2008a, Gr08) found a signal with strong significance and an amplitude reportedly higher than expected within the $\Lambda$CDM paradigm. We revisit the analysis using our own robust protocol, and extend the study to the two most recent and largest catalogues of voids publicly available. 
We quantify and subtract the level of foreground contamination in the stacked images and determine the contribution on the largest angular scales from the first multipoles of the CMB. We obtain the radial temperature and photometry profiles from the stacked images. Using a Monte Carlo approach, we computed  the statistical significance of the profiles for each catalogue and identified the angular scale at which the signal-to-noise ratio (S/N) is maximum. We essentially confirm the signal detection reported by Gr08, but for the other two catalogues, a rescaling of the voids to the same size on the stacked image is needed to find any significant signal (with a maximum at $\sim 2.4\,\sigma$). This procedure reveals that the photometry peaks at unexpectedly large angles in the case of the Gr08 voids, in contrast to voids from other catalogues. Conversely, the photometry profiles derived from the stacked voids of these other catalogues contain small central hot spots of uncertain origin. We also stress the importance of a posteriori selection effects that might arise when intending to increase the S/N, and we discuss the possible impact of void overlap and alignment effects. We argue that the interpretation in terms of an iSW effect of any detected signal via the stacking method is far from obvious.} 

\keywords{Cosmic background radiation - Dark energy - Large-scale structure of Universe}

\authorrunning{Ili\'c, S., Langer, M., Douspis, M.}
\titlerunning{On the detection of iSW with stacked voids}

\maketitle


\section{Introduction}
\label{sec:1}

Among the many sources of secondary anisotropies of the cosmic microwave background (CMB) temperature \citep[for a recent review,][]{2008RPPh...71f6902A}, the integrated Sachs-Wolfe \citep[henceforth iSW,][]{1967ApJ...147...73S} effect is particularly interesting in connexion with the acceleration of cosmic expansion. In principle, in a universe not dominated by cold matter, the energy of CMB photons is redshifted or blueshifted while they travel across cosmic over-densities or underdensities, respectively, owing to the stretching-out of the gravitational potentials created by the structures. Since the amplitude of this effect is expected to be rather small and shows mostly on the largest angular scales \citep[due to a line-of-sight cancellation,][]{1985SvAL...11..271K}, it has been suggested, first in the context of studies of the Rees-Sciama effect \citep[][]{Crittenden1996}, to cross-correlate the CMB temperature fluctuations with the distribution of tracers (e.g. galaxies) of gravitational potentials. After the first attempts to detect the iSW effect by that method \citep[][]{1998NewA....3..275B,2002PhRvL..88b1302B}, many studies have been published using the latest galaxy survey data. Often based on similar data and comparable techniques, the claims for detection range remarkably from a ``negligible'' to a $\sim 4.5\,\sigma$ statistical significance \citep[for a discussion, see][and \citealp{Giannantonio2012}]{Dupe2011}.

This puzzling situation calls for clarification. Discussions of the cross-correlation methods and techniques set aside, it would be invaluable to have another way to unveil the iSW effect. One such way would be to measure it directly in the CMB maps at the locations of the gravitational potentials that are responsible for it. Unfortunately, its amplitude with respect to the primary CMB anisotropies does not allow us to detect it structure by structure. However, stacking techniques can be profitably adopted to enhance the signal-to-noise ratio (S/N). With the help of the Wilkinson Microwave Anisotropy Probe (WMAP) five year maps \citep{2009ApJS..180..225H}, such a technique has been applied by \cite{Gran2008} to the supervoids and superclusters they identified in the catalogue of luminous red galaxies (LRGs) in the Data Release 6 (DR6) of the Sloan Digital Sky Survey (SDSS) \citep[see also][]{2008arXiv0805.2974G}. Their super-structure identification method uses the VOBOZ \citep[VOronoi BOund Zones,][]{2005MNRAS.356.1222N} and ZOBOV \citep[ZOnes Bordering On Voidness,][]{2008MNRAS.386.2101N} Voronoi tessellation-based, publicly available numerical codes. Focusing on the most significant (in terms of the density contrast) 50 supervoids and 50 superclusters, \citet{Gran2008} report a combined mean temperature deviation of 9.6 $\mu$K, at a significance just above $4\,\sigma$, which they interpret as a signature of the iSW effect. However, \citet{Carlos2012} pointed out that when considering several aperture scales (ranging from $1^\circ$ to $20^\circ$), a combined $\chi^2$ analysis yields a detection at a level of only $\sim 2\,\sigma$.

Using the Millennium simulation \citep{2005Natur.435..629S} and measuring the iSW effect that is expected in a standard $\Lambda$CDM universe, \cite{Gran2008} find that it is $\sim 2\,\sigma$ lower (at 4.2 $\mu$K) than what they obtained from the WMAP data. Other studies have also measured a somewhat higher  iSW effect than expected (e.g. \citealt{2008PhRvD..77l3520G}, \citeyear{Giannantonio2012}; \citealt{Ho2008}), although with small statistical significance. The high significance and the stronger-than-expected amplitude of the iSW effect detected through stacking have stimulated a number of investigations. For instance, \citet{Hunt2010} argue that LRGs need to be unnaturally \emph{underbiased} tracers of matter if we want to attribute the signal measured by \cite{Gran2008} to the iSW effect in the standard $\Lambda$CDM model. \citet{Nadathur2012} carefully analyse possible biasing effects arising from the strategy adopted by \cite{Gran2008}, but show that even when those selection effects are taken into account, the signal from supervoids expected within standard $\Lambda$CDM is still at odds by $> 3\,\sigma$ with the value measured by \citet[][for similar conclusions reached with numerical simulations of the iSW effect, see \citealt{Flender2012}]{Gran2008}. On the other hand, \citet{Papai2010a} describe the superstructures by uncompensated density profiles from  Gaussian statistics and find that large temperature deviations, $\Delta T \sim 10 \mu$K, can be obtained. Moreover, building on this result, \citet{Papai2011} re-assessed  the statistical significance of those values and estimated that the discrepancy between the observations and the theoretical $\Lambda$CDM  predictions is  $2\,\sigma$. However, as noted in \citet{Nadathur2012}, the latter interpretation would imply voids with unphysical underdensities $\delta < -1$. Finally, we note that \citet{Granett2009} analysed again the WMAP 5 CMB data with a template fitting method and confirmed the $4\,\sigma$ significance of the signal found previously. Furthermore, they also reconstructed an iSW map from the very same LRG sample they used for their previous superstructure identification. On that reconstructed map, however, the combined voids and clusters associated temperature deviation was only $0.08 \pm 0.1 \mu$K, casting doubt on the suggestion that the signal is indeed due to the linear iSW effect.

Should we conclude that the large CMB temperature deviations measured in association with superstructures signal a tension with the $\Lambda$CDM model? In this paper, we would like to take a further step towards answering that question. Since the study of \cite{Gran2008}, new CMB maps have been released and other superstructure catalogues have been published. We do the stacking analysis with the new data and look for the iSW signal that could be associated with the large scale structure. We pay particular attention to the bias introduced into the results by selection effects and illustrate it with an explicit example.
We begin by introducing the data used for this study. In Sect.~\ref{sec:3}, we outline the methodology we adopted, stressing  the assessment of the statistical significance of our results. The latter are detailed in Sect.~\ref{sec:4}, and in Sect.~\ref{sec:5} we discuss the importance of a posteriori selection effects that may artificially boost the statistical significance, as well as other sources of uncertainty that cloud the interpretation of the results. Finally, we conclude in Sect.~\ref{sec:CCL}. We use the parameters from WMAP 7 \citep{komatsu2010} best-fit cosmology for all relevant calculations.

\section{The data}
\label{sec:2}

\subsection{Cosmic microwave background}
\label{sec:2-1}

In the present study, we have used the maps of the cosmic microwave background released by the WMAP team after seven years of observation \citep{Jarosik2011}, in contrast to the five-year data used by \citet[][Gr08 thereafter]{Gran2008} in their study. We took the individual channel maps at the three frequencies that are the less contaminated by foregrounds (the Q, V, and W bands at 41, 61, and 94~GHz, respectively). We also used the hit map of the WMAP mission, i.e. the map that contains for each pixel the number of times it was observed in the satellite lifetime. The impact of foregrounds and the associated possibility of false signals is often a source of uncertainty in iSW studies. We therefore consider and assess their possible influence by redoing our analyses on the foreground reduced maps released by the WMAP team in the same frequency channels.\footnote{All maps can be downloaded from the \textsc{lambda} website \\ \url{http://lambda.gsfc.nasa.gov}}

\subsection{Granett et al. (2008)}
\label{sec:2-2}
The first catalogue of superstructures (clusters and voids) that we considered was created and studied by Gr08. Since it was already explored with WMAP 5-year data, it will serve us as a ``fiducial" set when testing all the steps of our own stacking procedure. This will also be the opportunity to revisit the work of Gr08 with the newer seven-year data from WMAP.

The catalogue is based on the LRG sample of the SDSS DR6  \citep{Adelman2008}, which is composed of 1.1 million LRGs in the range $0.4 < z < 0.75$ (with a median $z = 0.52$), covers 7500 square degrees on the sky, and occupies a volume of $\sim 5\ h^{-3} \mathrm{Gpc^3}$. In this survey, Gr08 searched for clusters and voids respectively using  the two publicly-available structure-finding algorithms VOBOZ and ZOBOV based on Voronoi tessellation. They detected 631 voids and 2836 clusters above a $2\,\sigma$ ``significance level", defined as the probability of obtaining the same density contrasts as those of voids and clusters  in a uniform Poisson point sample.

From these results, they kept and released the 50 most significant clusters and 50 voids only, which they used in their CMB stacking analysis. The catalogue of these 100 superstructures contains all the information needed for their analysis. For each structure it provides the position of its centre on the celestial sphere, the mean and maximum angular distances on the sky between the galaxies in the structure and its centre, its physical volume, and three different measures of its density contrast (either calculated from all its Voronoi cells, from only its over/underdense cells, or from only its most over/underdense one).

After conversion to physical distances (see details in Sect.~\ref{sec:2-5}), we note that these voids have a mean effective radius of about 78 Mpc and a mean redshift of $\sim0.5$, corresponding to angular sizes on the sky of about 3.5 degrees.
1911

\subsection{Pan et al. (2012) void catalogue}
\label{sec:2-3}
\cite{Pan2012} published a catalogue of cosmic voids and void galaxies identified in the seventh data release (DR7) of the SDSS. Using the VoidFinder algorithm as described by \cite{Hoyle2002}, they identified and catalogued 1055 voids with redshifts lower than $z=0.1$. For each void they provide its position on the sky (also, but not useful for us, its 3D position in the survey), its physical radius (defined as the radius of the maximal sphere enclosing the void), an effective radius (as voids are often found to be elliptical) defined as the radius of a sphere of the same volume, its physical distance to us, its volume, and mean density contrast.

The filling factor of the voids in the sample volume is 62\%. The largest void is just over 47 Mpc in effective radius, while their median effective radius is about 25 Mpc. Some of them are both very close to us and relatively large (more than 30 Mpc in radius) resulting in large angular sizes on the sky, up to 15 degrees and above.
 993

\subsection{Sutter et al. (2012) void catalogue}
\label{sec:2-4}
The most recent catalogue considered in the present study was released by \cite{Sutter2012}.\footnote{Catalogue published on-line at \\ \url{http://www.cosmicvoids.net}, version 21/02/2013.} Using their own modified version of the void finding algorithm ZOBOV, Sutter et al. also built a void catalogue from the SDSS DR7, taking particular care to account for the effects of the survey boundary and masks. 
In the latest version of their catalogue, they found a total of 1495 voids, which they divided into six distinct subsamples of increasing redshift: four from the main SDSS  (named \textit{dim1}, \textit{dim2}, \textit{bright1}, and \textit{bright2}) and two from the SDSS LRG sample (\textit{lrgdim} and \textit{lrgbright}). The redshifts of these voids span $z\sim[0,0.4]$, while their sizes range approximatively from 5 to 150 Mpc. A summary of the six subsamples and their contents is provided in Table~\ref{tab:Sutter}, together with the contents of the other two catalogues described in Sects.~\ref{sec:2-2} and \ref{sec:2-3}.

This catalogue stands out by the amount of information provided about its voids: position of the barycentre, redshift, effective radius (with the same definition as in the Pan et al. catalogue), locations of member galaxies, one-dimensional radial profiles of stacked voids, two-dimensional projections of stacked voids, and other statistical information about their distribution. Since its first release in July 2009, the catalogue has been subjected to regular updates and modifications reflecting improvements in the detection algorithm and bug corrections, along with inclusion of additional void data. 

\begin{table}[t]
	\caption{Summary of the void catalogues.}
	\centering
	\begin{tabular}{ c c c }
		\hline
		\hline
		\textbf{(Sub)sample name} & \textbf{Redshift range} & \textbf{Number of voids} \\
		\hline
		\textbf{Sutter et al.} & $0.003<z<0.43$ & \textbf{1495} \\
		\hline
		dim1  & $0.003<z<0.048$ & 218 \\ 
		dim2 & $0.05<z<0.1$ &  419 \\
		bright1 & $0.1<z<0.15$ & 341 \\
		bright2 & $0.15<z<0.20$ & 176 \\
		lrgdim & $0.16<z<0.35$ & 291 \\
		lrgbright & $0.36<z<0.43$ & 50 	\\
		\hline
		\textbf{Granett et al.} & $0.43<z<0.69$ & \textbf{50} \\
		\hline
		\textbf{Pan et al.} & $0.009<z<0.1$ & \textbf{1055} \\
		\hline
	\end{tabular}
	\label{tab:Sutter} 
\end{table}

\begin{figure*}[!ht]
\centering
\includegraphics[width=0.8\textwidth]{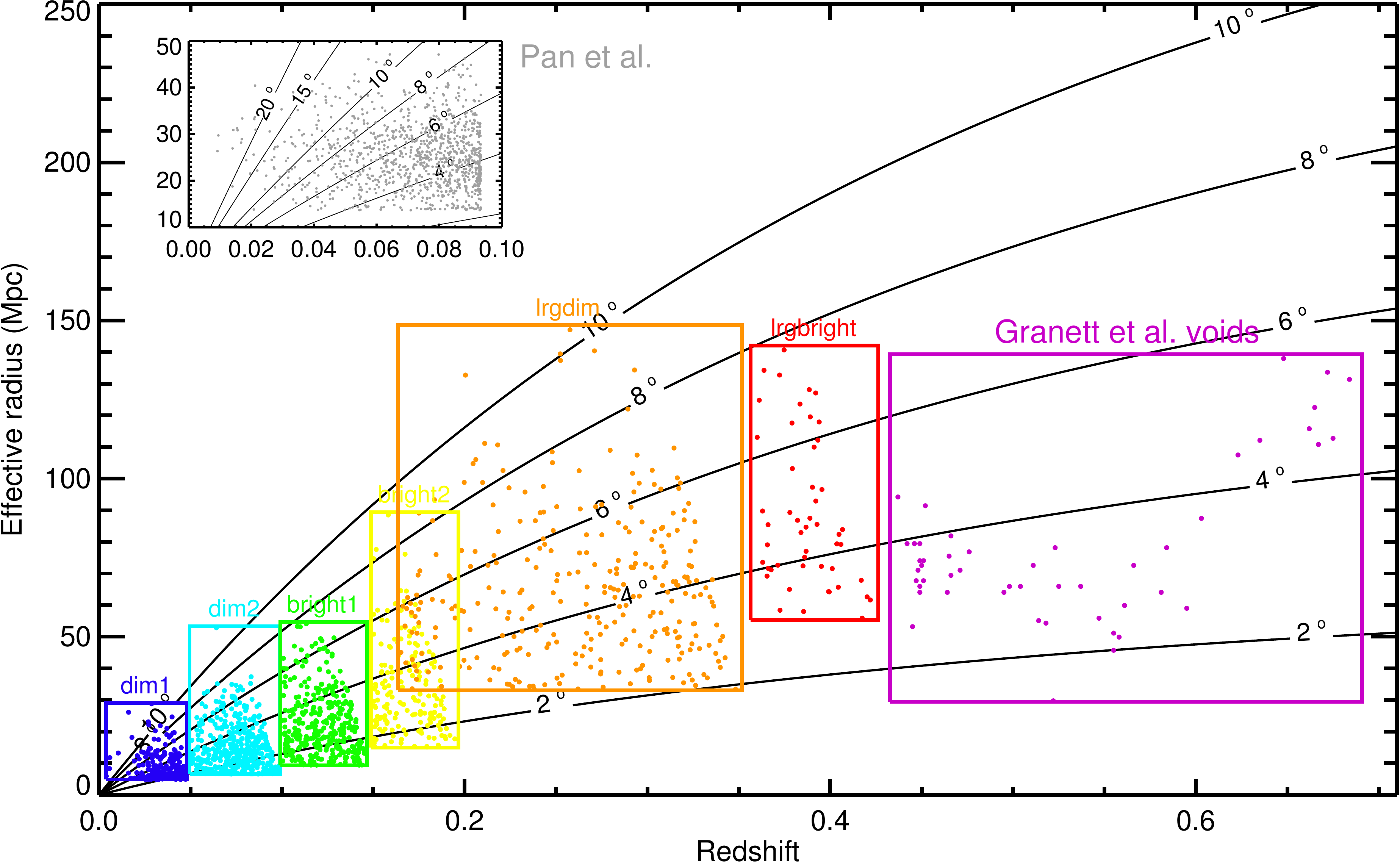}
\caption{\small Effective radii (in Mpc) as a function of redshift for all the voids used in this work. The two catalogues of Granett et al. and Pan et al, and the six subsamples of Sutter et al. are delimited by boxes labelled by their names. The black curves are here to indicate the angular size (in degrees) of a structure in the redshift-radius plane.}
\label{fig:voids:all}
\end{figure*}
2985

\subsection{Using the void catalogues}
\label{sec:2-5}
The only indispensable information that we extract from these catalogues is the positions on the sky of the observed structures (most frequently the position of their barycentre), which is essential to apply the stacking procedure. They are most often provided in celestial coordinates (RA, DEC) that we easily convert into galactic coordinates in order to use them in the framework of the HEALPix\footnote{\url{http://healpix.jpl.nasa.gov}} suite. For more advanced analysis and interpretations, we first require the redshift of these structures, which is a parameter that is directly given in the considered catalogues, except for that of Pan et al., which we obtained from the physical distance to the voids. Finally we also make use of the physical radius of the structures, through its relation to the angular size on the sky (see Sect.~\ref{sec:5-2}). When provided, we use the effective radius of the structure (Pan et al., Sutter et al.), which we translate into an angular size using the available redshift information. For the Gr08 catalogue, we derive the effective radius -- with the same definition as in the two other catalogues -- from the provided volume, and then also convert it to an angular size. This is somewhat of a compromise between the two angular sizes already provided by Gr08: the \textit{mean radius} between the centre of the void and all its Voronoi cells (possibly an underestimate of the void size on the sky), and the maximum radius between the centre and the farthest cell (possibly an overestimate). 

Of the Gr08 catalogue, we only use the voids and disregard the clusters, for the sake of consistency with the other two catalogues, but also because voids are more adapted to stacking studies; indeed, emission of an astrophysical origin is less likely to contaminate the iSW signal from these objects. As a summary, we plot the properties of the voids from all catalogues in Fig.~\ref{fig:voids:all}, including their redshifts, effective radii, and corresponding angular sizes.2016

\section{Methodology}
\label{sec:3}
\subsection{Initial procedure}
\label{sec:3-1}

The analysis of these three different catalogues requires us to have a robust and well defined procedure for a systematic analysis of all the structures considered. First, the standard stacking procedure that we apply in this study consists in the following steps, for each superstructure sample or subsample:
\begin{itemize}
	\item We first take a CMB map of the WMAP data at a given frequency, either raw or foreground cleaned, from which we remove the cosmological monopole and dipole;
	\item We construct the associated weight map by taking a galactic mask (here the KQ75 from the WMAP team, the extended mask for temperature analysis that removes $\sim 22\%$ of the sky along the galactic plane and around point sources) and multiplying it by the hit map associated with the survey;
	\item We retrieve the galactic longitudes and latitudes of the structures we study;
	\item We use a custom code based on the HEALPix package to cut a patch in the CMB map centred on each structure. We choose the patches to have a 6 arcmin/pixel resolution (small enough to oversample any of the CMB maps used) and to be squares of $301\times301$ pixels, i.e.  $30^\circ\times30^\circ$ patches;
	\item At the same time, we cut identical square patches at the same locations in the associated weight map;
	\item The final stacked image is then constructed as the average image of all CMB patches weighted by their corresponding weight patches.
\end{itemize}

\noindent Two main products are then extracted from the stacked image:
\begin{itemize}
	\item the radial temperature profile starting from the centre of the image, by computing the mean temperature of the pixels in rings of fixed width and increasing angular radius. Considering the characteristics of our stacked images, it is calculated here for 150 radii between $0^\circ$ and $15^\circ$, with a width of $\Delta\theta=0.1^\circ$;
	\item the aperture photometry profile, using a compensated filter approach. At each angle $\theta$, we compute the photo\-metry as the difference between the mean temperature of the pixels inside the disk of angular radius $\theta$ and the temperature of the pixels in the surrounding ring of same area, i.e. the ring enclosed between circles of radii $\theta$ and $\theta\sqrt{2}$. With this procedure, we obtained this profile for 150 angles between $0^\circ$ and $(15/\sqrt{2})\sim10.6^\circ$.
\end{itemize}
The summation of square pixels contained inside a disk can lead to calculation errors due to omitted fractions of pixels close to the boundaries of the disk. To reduce these as much as possible, we upscale the $301\times301$ stacked image into a $1204\times1204$ one -- each pixel of the original image is divided into 16 sub-pixels of the same value. Statistical errors for these two profiles are estimated by computing the standard deviation of each calculated mean of pixels. We mainly focus on the analysis of these two profiles (the image itself is useful for illustration purposes only), where we look for any remarkable signal whose significance we assess below (see Sect.~\ref{sec:3-3}).

We note that the size of the CMB patches is chosen large enough to enclose any of the structures studied and to properly compute their photometry (based on the effective angular radius that has to be smaller than $\sim 10.6^\circ$), with the exception of 11 voids (out of 1495) in the Sutter et al. catalogue and 21 in the Pan et al. one (out of 1055); we omit these voids in the stacking.

\subsection{Choice of maps}
\label{sec:3-2}

Each of the three CMB maps that we use (WMAP Q, V, and W) inherently has a different resolution and contains different types and levels of foregrounds that may contaminate them. Before progressing any further, we assess the impact of the properties of each map, using our fiducial stacking (i.e. using the Gr08 void positions) as a basis.

\subsubsection{Effects of resolution}
\label{sec:3-2-1}

\begin{figure}[t]
\centering
\includegraphics[width=0.35\textwidth]{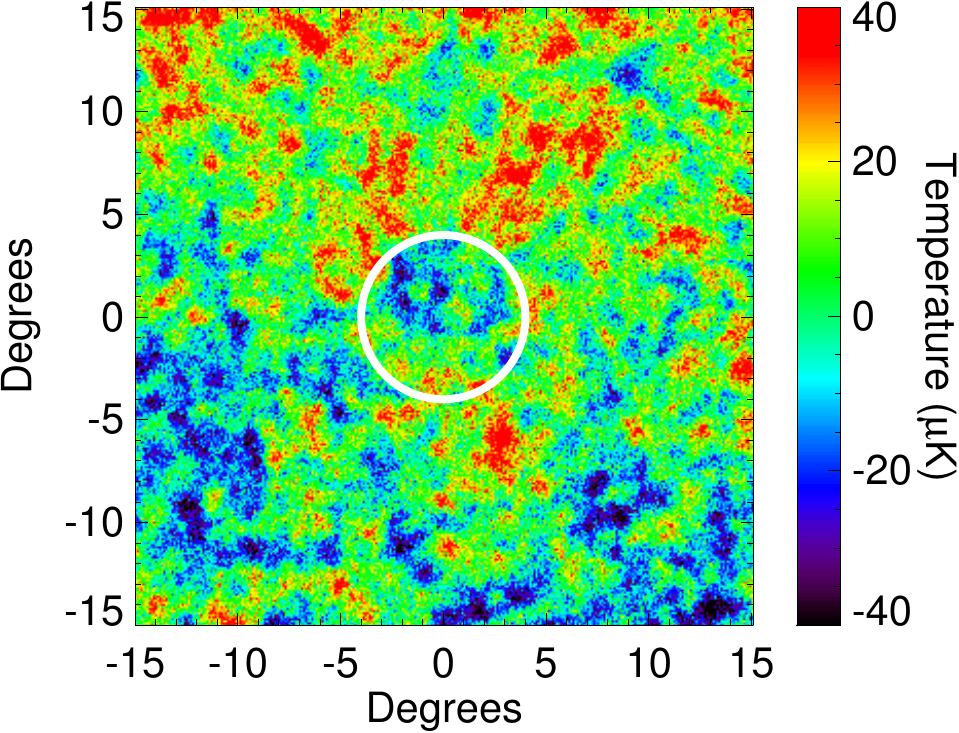}
\caption{\small Image resulting from our stacking procedure at the location of the 50 voids of Gr08, here using the V band CMB map of WMAP. A cold spot, reportedly due to the iSW effect, is visible roughly at the centre of the image with an angular radius of about~$4^\circ$  (underlined by the $4^\circ$ radius white circle centred on the image).}
\label{fig:stack}
\end{figure}

\begin{figure}[t]
\centering
\includegraphics[width=0.45\textwidth]{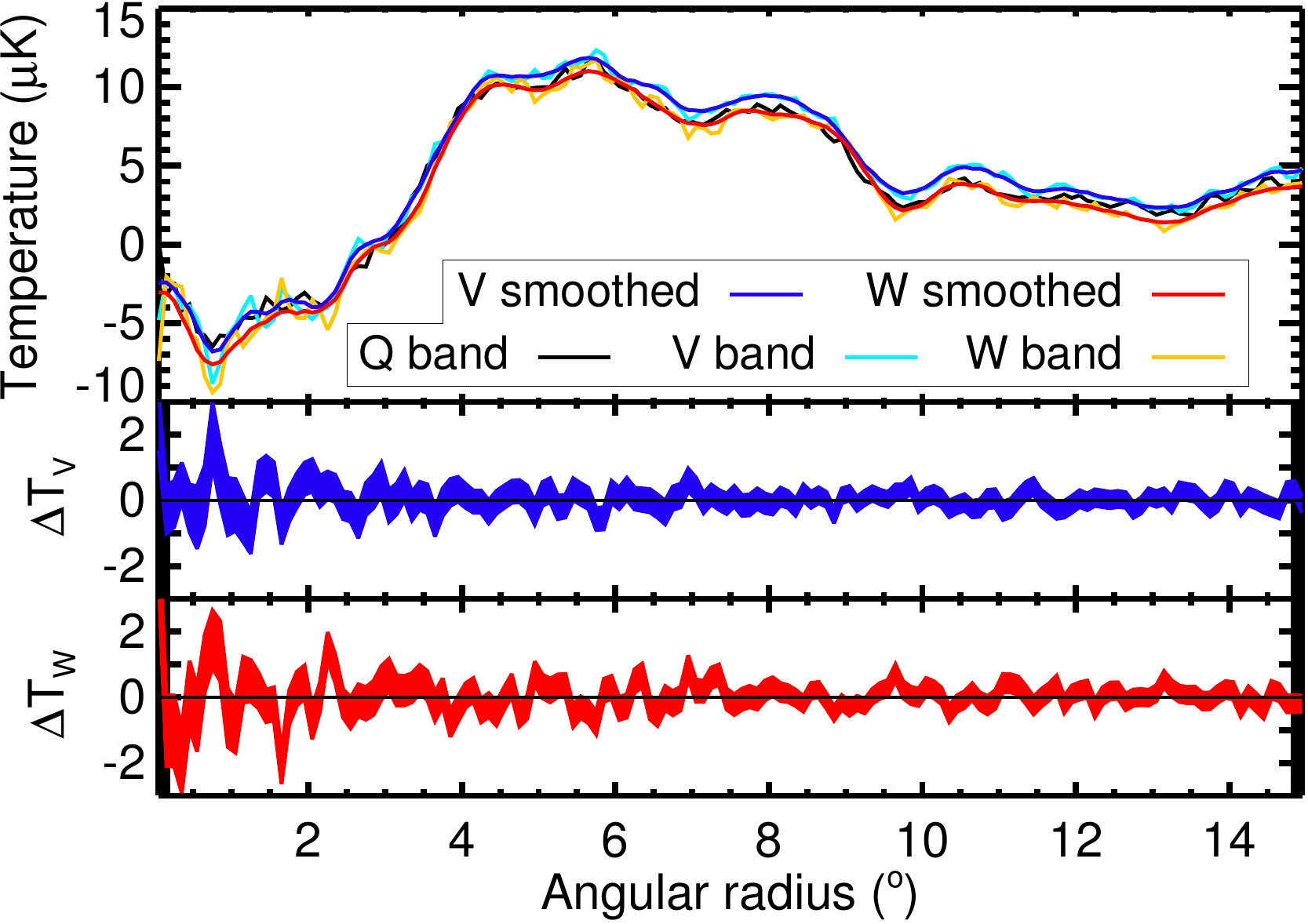} \\ \ \\ \includegraphics[width=0.45\textwidth]{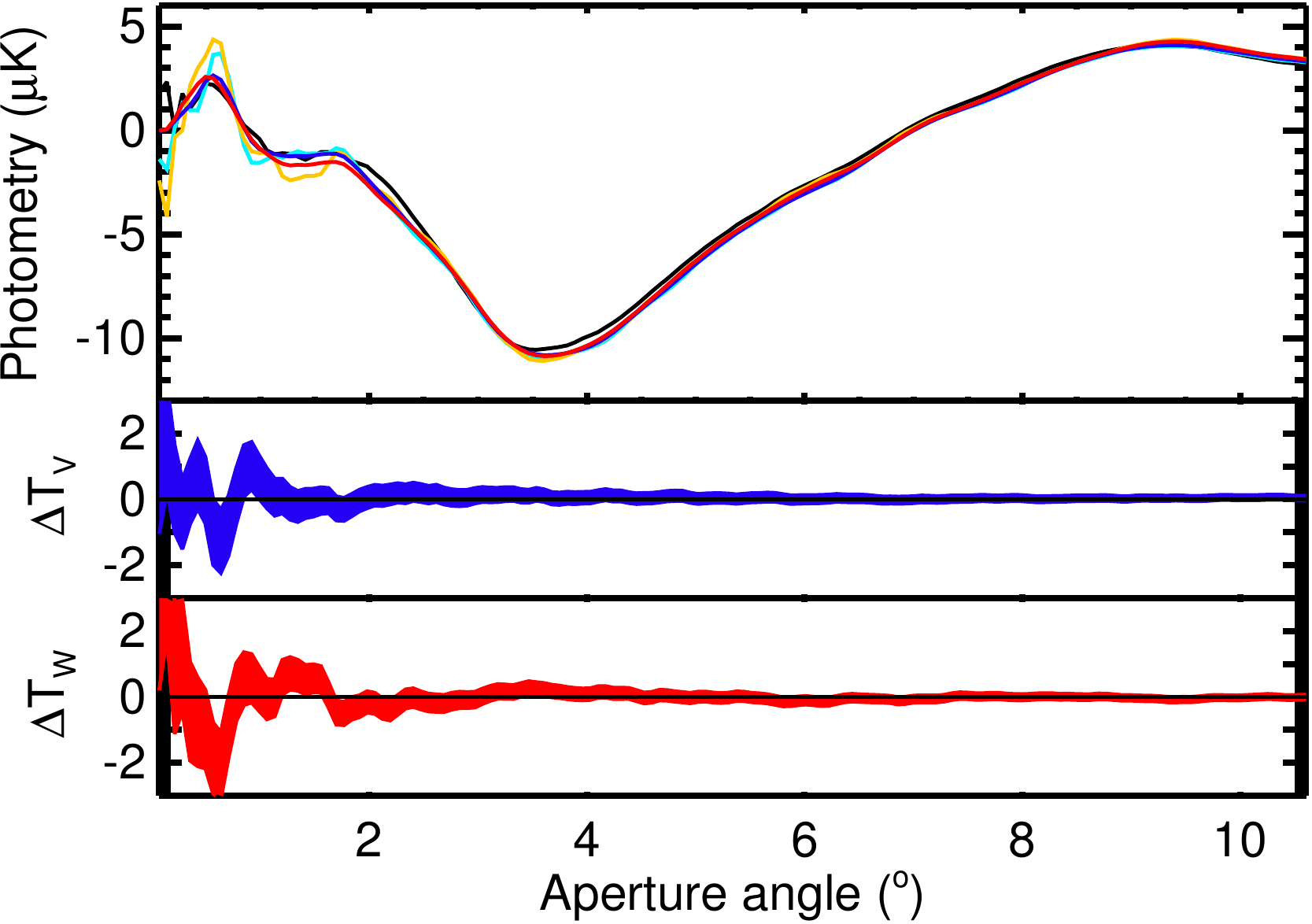}
\caption{\small \textit{First plot}: radial temperature profiles (\textit{top panel}) of the stacking of Gr08 voids, done on WMAP Q, V, and W maps (both in native resolution and smoothed by a 30.6 arcminutes kernel). The differences in the profiles between the smoothed and original maps are plotted below the main plot (\textit{middle}: V band; \textit{bottom}: W band). The width of the shaded curves corresponds to the statistical errors on the profile measurements. \textit{Second plot}: Same graphs and legend as above but for the aperture photometry profiles of the stacking of Gr08 voids.}
\label{fig:smooth}
\end{figure}

We perform the fiducial stacking of the Gr08 voids on the three raw CMB maps from the Q, V, and W frequency bands, the beam sizes of which are, respectively, 30.6, 21 and 13.2 arcminutes ; an example of a stacked image is shown in Fig.~\ref{fig:stack}. To have a consistent stacking analysis through all the considered frequencies, we need to ``standardise" those maps by first smoothing them at the lowest resolution of the three (the Q band map) in order to lose as little information as possible. We then redo the fiducial analysis, and the resulting profiles are plotted in Fig.~\ref{fig:smooth}.

The stacks at each frequency, both raw and smoothed, give roughly the same results with only percent-level differences especially for the photometry profiles -- the most useful products here. The degradation of the V and W maps to the lower resolution of the Q map naturally smooths the measured profiles and reduces their dispersion around the results of the Q band, which is the desired effect. Otherwise, this procedure does not significantly modify their amplitude and angular dependence, so that we may adopt hereafter this new common resolution for all frequencies. 

In all cases, a signal appears in the photometry profile with a maximum (in absolute value) on an angular scale of about $3.5^\circ$, the preferred size changing only very slightly between frequencies. We keep in mind that the smoothing procedure with a $\sim30.6$ arcminute beam blurs the information and details contained below this scale, therefore we should not devote too much attention to any feature in the profiles at angles lower than this value.

\subsubsection{Assessment of the effects of the foregrounds}
\label{sec:3-2-2}

One other source of concern comes from the influence of foregrounds present in the CMB maps, because they might mimic the expected iSW signal in the stacked images. To assess their possible impact, we performed the stacking of the Gr08 voids first on raw and then on foreground cleaned CMB maps at all frequencies. We then look for differences between the two results, either in the amplitude or in the shape of the signal.   

\begin{figure}[t]
\centering
\includegraphics[width=0.45\textwidth]{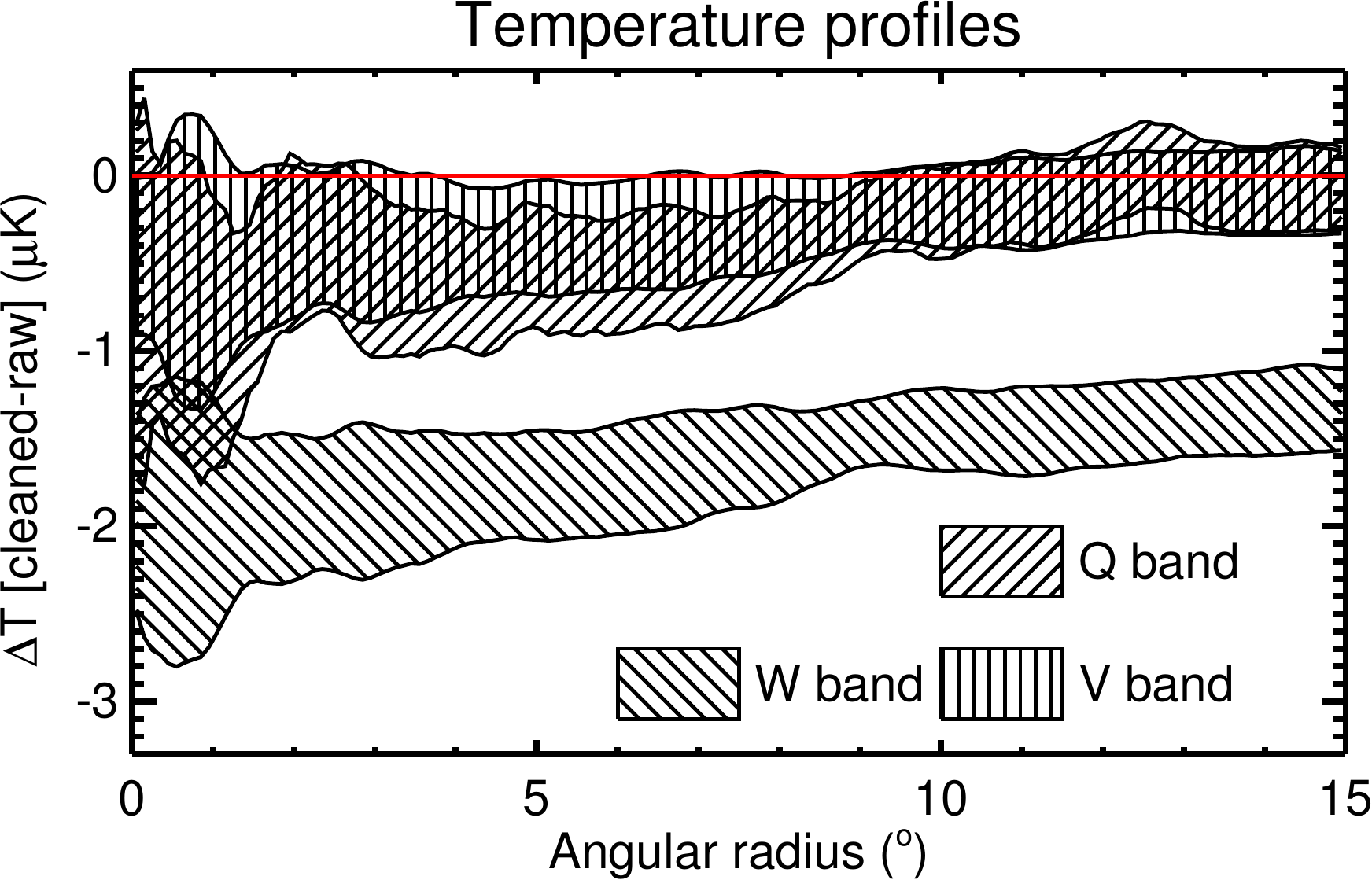} \\ \ \\ \includegraphics[width=0.45\textwidth]{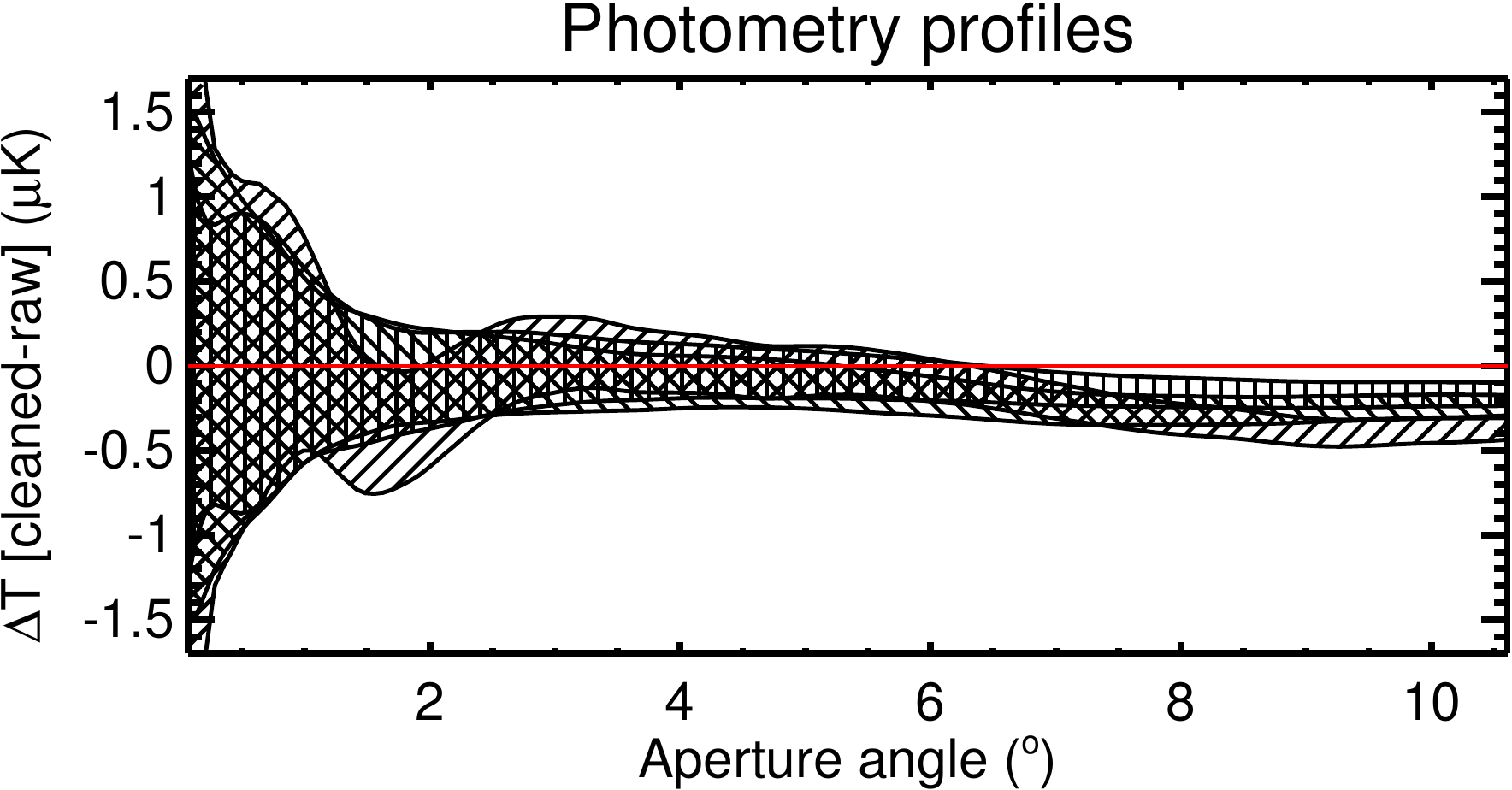}
\caption{\small For the Gr08 stacking, differences in the temperature (\textit{top}) and photometry (\textit{bottom}) profiles between foreground-cleaned maps and raw maps, for the three frequency bands considered. The width of the shaded curves corresponds to statistical errors on the profile measurements. The quasi-flat offsets observed in the temperature profiles do not affect the photometry substantially.}
\label{fig:fore}
\end{figure}

Results are illustrated in Fig.~\ref{fig:fore}: we obtain systematic offsets of a few micro-kelvins in the radial temperature profiles and less in the photometry profiles. This indicates that we mainly remove an almost uniform background, which does not influence  the aperture photometry of the stacked image much. As a precaution we then use the foreground cleaned maps for our analyses. Possible residuals in the cleaned maps should not be of any concern since they would have the same ``flat" behaviour in the stacked images. 

\subsubsection{Analysis of the temperature gradient}
\label{sec:3-2-3}

\begin{figure*}[!ht]
\centering
\includegraphics[height=4.cm]{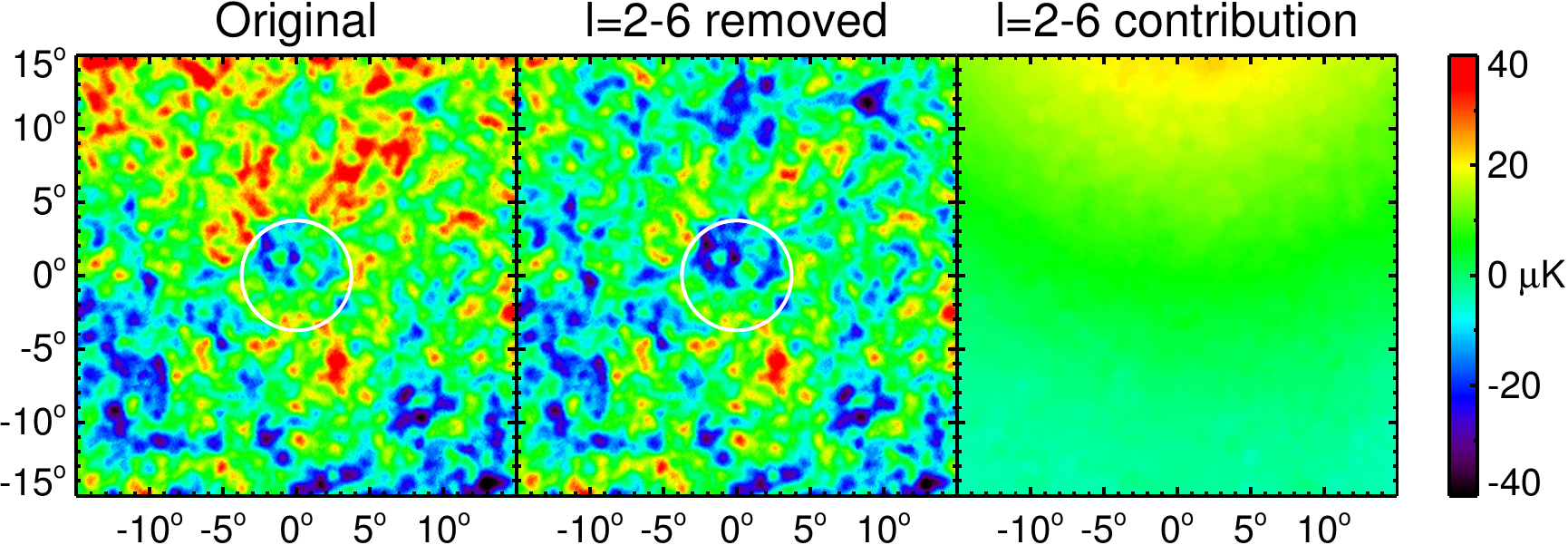} \hspace{0.5cm} \raisebox{-0.5cm}{\includegraphics[height=4.5cm]{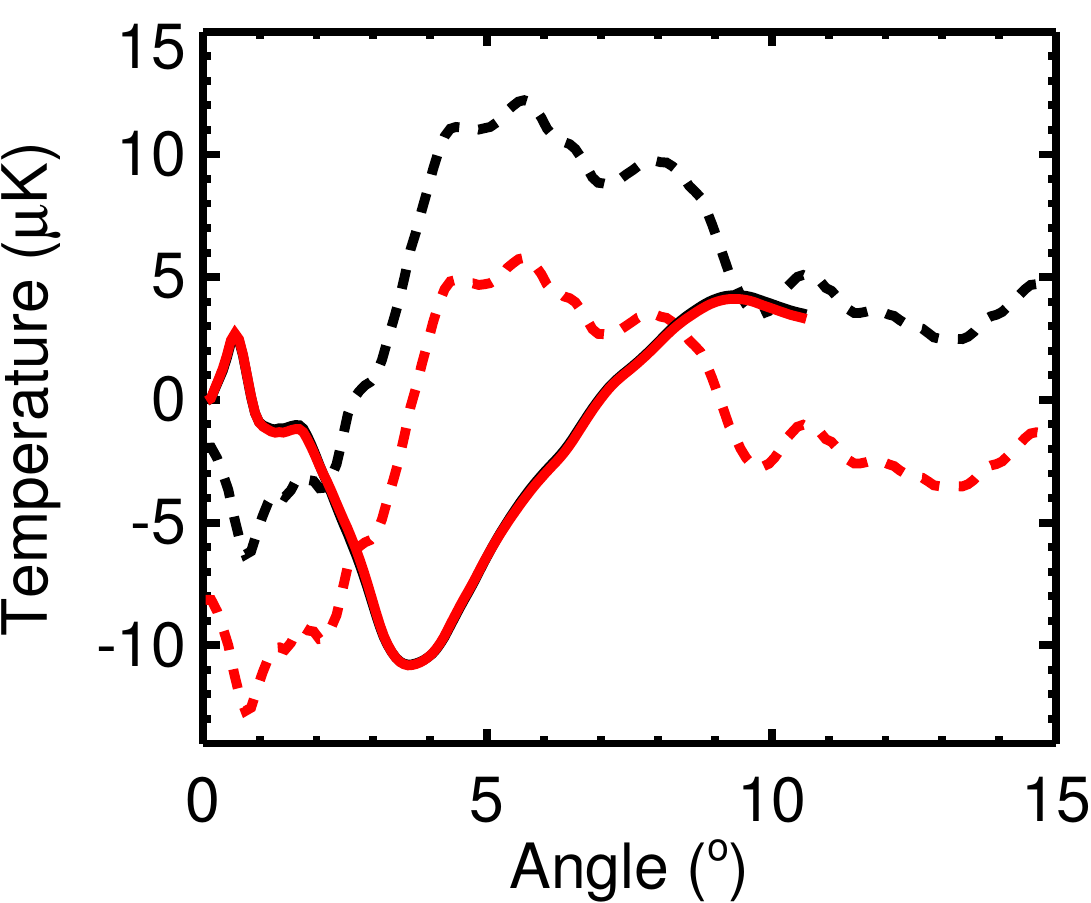}}
\caption{\small Images representing the stacking of the Gr08 voids done (from left to right) on the cleaned WMAP V map, on the same map without its $\ell=2-6$ multipoles, and on these multipoles only. The temperature (dotted curves) and photometry (solid) profiles shown in the rightmost plot are obtained from the first (black curves) and second (red) stacked images. The temperature offset induced by the removed multipoles does not affect the photometry.}
\label{fig:poles}
\end{figure*}

Another map-related problem showed up during our investigations, when we observed that a clear temperature gradient appeared in many of our stacked images with the new catalogues, roughly on a north-south axis with hotter high latitudes (see an example in Fig.~\ref{fig:poles}). Foregrounds can be excluded as a possible source of this,  because the gradient appears in both raw and foreground-cleaned maps, and also to a lesser extent because it is contrary to our intuition about foreground contamination since we would expect hotter temperatures closer to the galactic plane. 
One may wonder why this gradient was not observed by Gr08, but this can be explained by their rotating the CMB patches before stacking them in order to align the major axes of the voids. Since the voids are randomly oriented  a priori, introducing such rotations is enough to erase any systematic gradient. In our study, we did not have any access to the information on the orientation of the voids (it is not included in the Gr08 catalogue) so we did not rotate the CMB patches. A systematic gradient in the final image is clearly observed (see Figs.~\ref{fig:stack} and \ref{fig:poles}).  

\begin{figure}[t]
\centering
\includegraphics[width=0.4\textwidth]{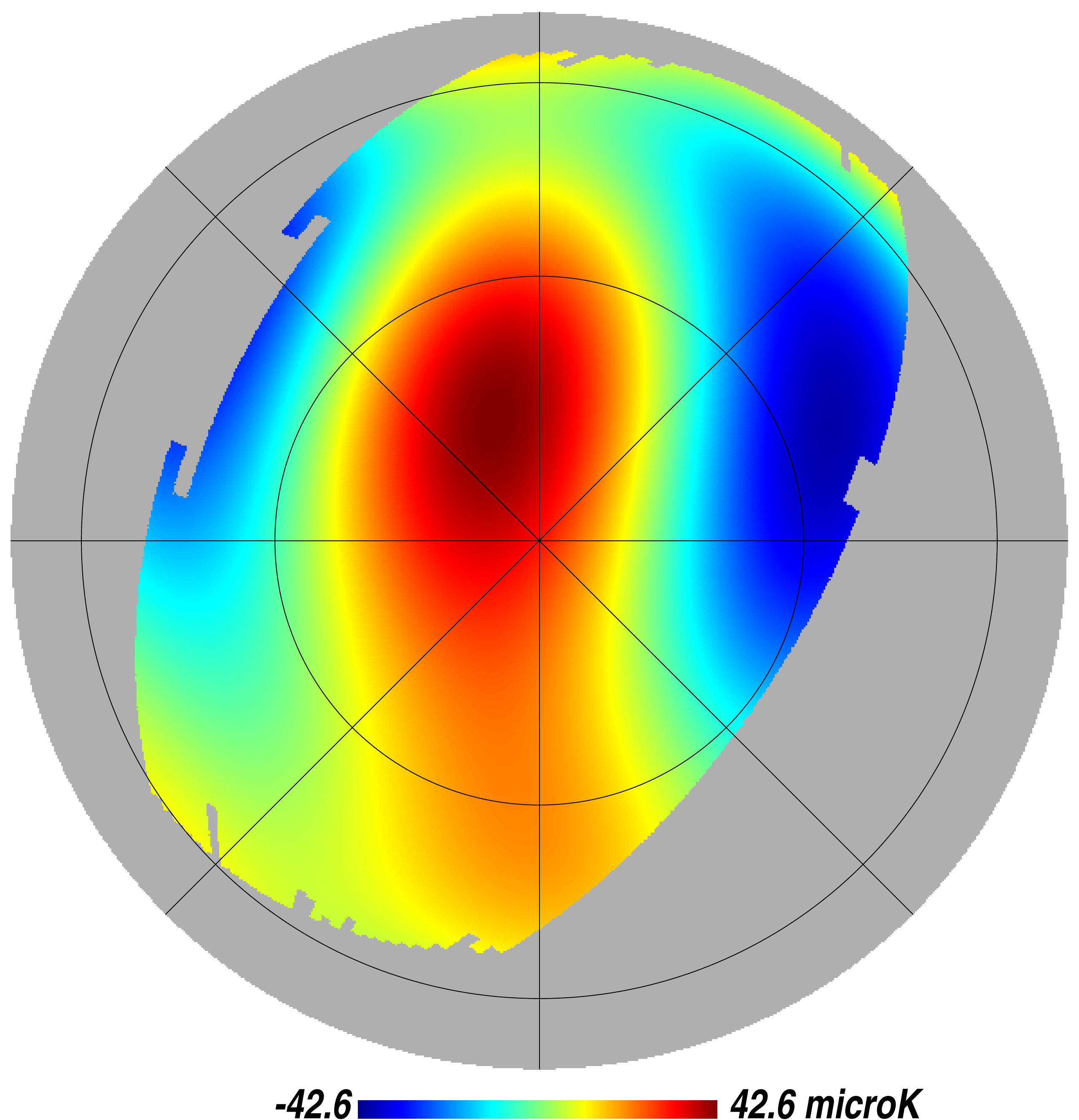}
\caption{\small Orthographic projection of the $\ell=2$ to $6$ multipoles map extracted from the foreground reduced WMAP Q map, in galactic coordinates. Only half of the map is visible (centred on the galactic north pole), with a mask showing only the area covered by the main SDSS. A graticule grid has been superposed with a $45^\circ$ step in longitude and $30^\circ$ in latitude.}
\label{fig:polmap}
\end{figure}

After discarding foregrounds as culprits, we looked for ``intrinsic" causes of this gradient. This led us to decompose the CMB maps on the spherical harmonics and to analyse the individual contribution of each multipole. This approach proved  fruitful since it appears that the measured gradient is mainly caused by the first few multipoles of the CMB maps we use, especially by the $\ell=6$ multipole map. In the region of the sky covered by the SDSS (where all the superstructures we considered are located), these multipoles combine to yield indeed a strong north-south gradient (see Fig.~\ref{fig:polmap}), which will be present at some level in every patch of CMB, hence in the final stacked image.

Visually speaking, subtracting the contribution of these multipoles out of the CMB maps does effectively remove the gradient in the stacked image. But on the other hand, the effect is almost negligible in the photometry profile of the Gr08 voids: indeed, the removed contribution  most often has the shape of a simple tilted plane (see Fig.~\ref{fig:poles} for an example), which does not affect the aperture photometry since it is equivalent to a constant background in the calculation. Indeed this gradient affects very low multipoles ($\ell\sim2$-6) that correspond to angular scales larger than $\sim30^\circ$, well above the angular sizes considered in the stacked profiles ($\theta<15^\circ$). For this reason, we will keep using the original CMB maps in our studies, which will also prevent the possible removal of relevant information in the stacked images. However since the other two catalogues that we use contain voids with larger angular sizes, we should keep the influence of this gradient in mind in the remainder of the analysis. We should also remember that all the temperature profiles will be affected by a systematic offset as shown in Fig.~\ref{fig:poles}.

\subsection{Significance estimation}
\label{sec:3-3}

When taken alone, the stacked images and their associated profiles are not enough to conclude anything about a possible detection of the iSW effect. Any peculiar feature that seems to stand out could very well be a random event well within statistical fluctuations. As a consequence, we have to take  great care in assessing the significance of our results. 

We have devised a systematic way to compute the significance adopting a Monte Carlo approach. We consider the stack of $N_v$ voids, identified in the data, whose significance we try to estimate. We pick many sets (at least 10000) of $N_v$ random positions on the sky confined within the area covered by the SDSS. For each random set, we perform the same analyses as for the data voids; i.e., we produce a stacked image of the $N_v$ patches extracted from the same CMB maps, and compute its radial temperature and photometry profiles. We store all these profiles in memory and end up with thousands of temperature profiles (called here $T^i_\mathrm{sim}(\theta)$) and photometry profiles ($P^i_\mathrm{sim}(\theta)$).
After this, for each angular size of the profiles, we compare the results from the stack of data voids to the statistical distribution of results from the random stacks. In practical terms, we calculate the S/N of the data temperature ($T_\mathrm{data}(\theta)$) or photometry ($P_\mathrm{data}(\theta)$) profiles, at each angle $\theta$ considered, as follows:
\begin{equation}
	\label{eq:SNR}
	\mathrm{S/N}_{T,P}(\theta) = \frac{\left|(T,P)_\mathrm{data}(\theta)  - \mathrm{Avg}[(T^i,P^i)_\mathrm{sim}(\theta)]\right|}{\mathrm{StdDev}[(T^i,P^i)_\mathrm{sim}(\theta)]}
\end{equation}
\noindent where the average and the standard deviation are evaluated over the collection of random stacks. We then obtain two S/N angular profiles for the considered stack: one of its temperature and one of its photometry.    

\begin{figure}[t]
\centering
\includegraphics[width=0.45\textwidth]{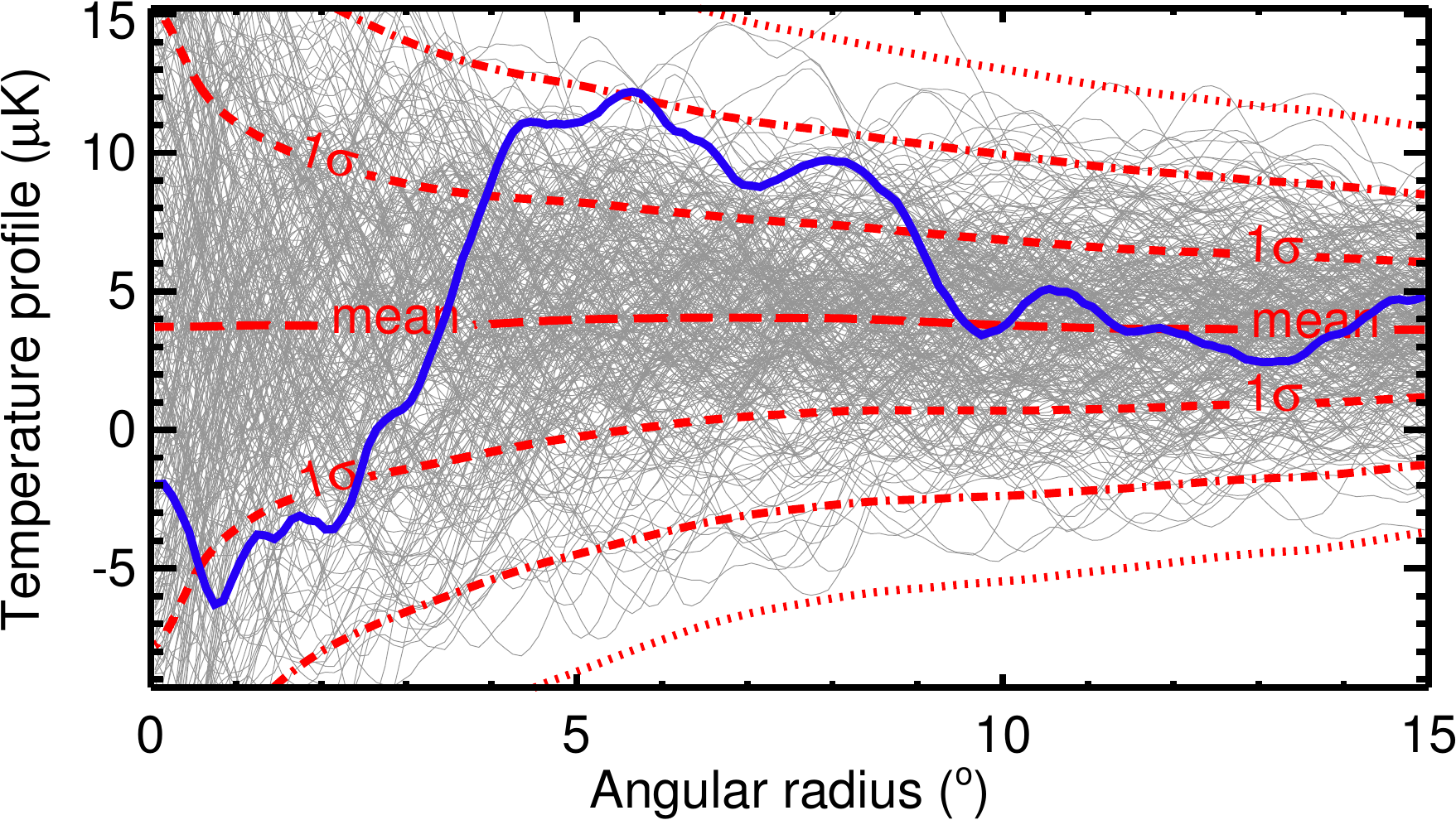} \\ \vspace{0.2cm} \includegraphics[width=0.45\textwidth]{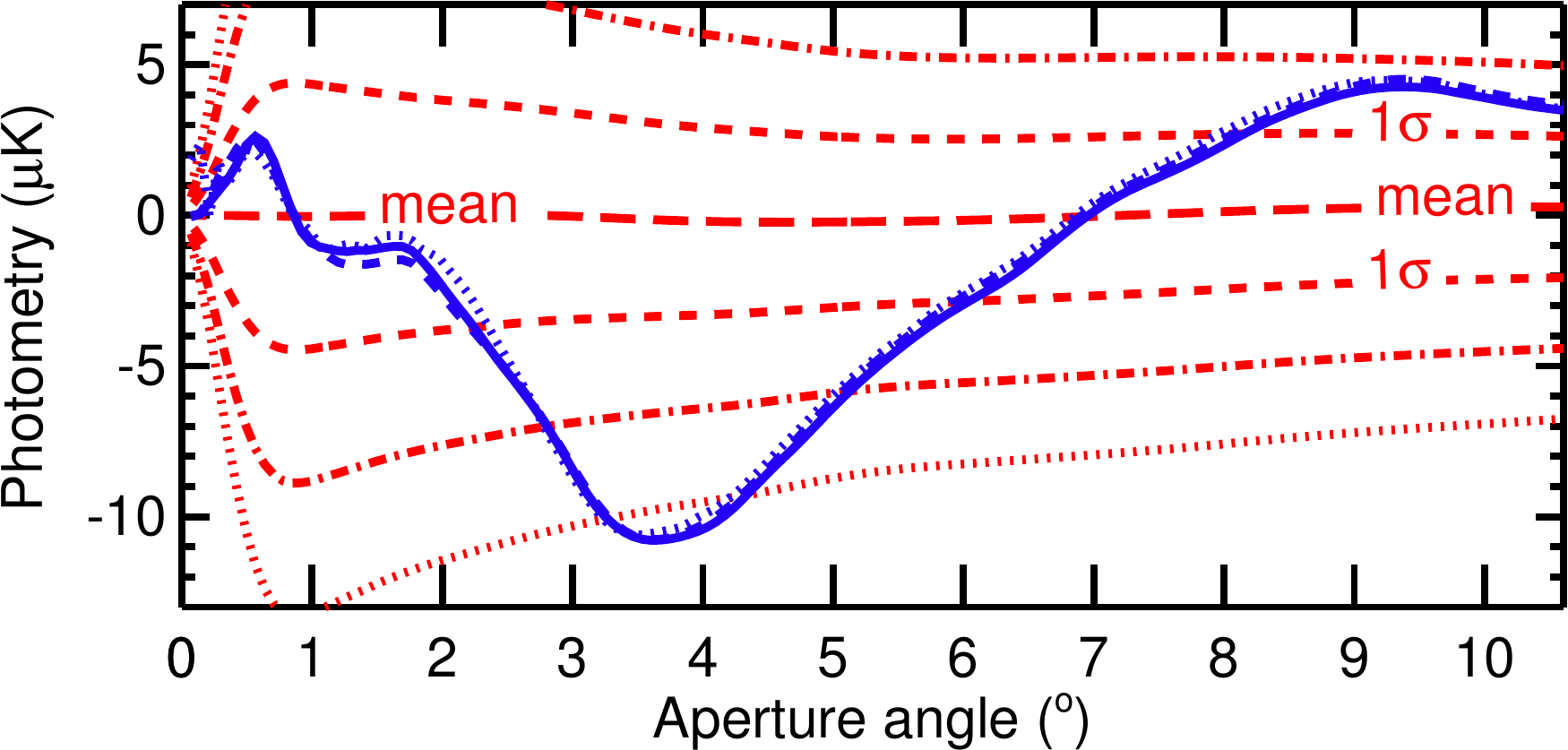}
\caption{\small \textit{Top panel:} Mean profile (long red dashes) from all 14000 random stacks and the $1\,\sigma$ (dashed), $2\,\sigma$ (dash-dotted), and $3\,\sigma$ (dotted) limits of the distribution of profiles. For illustration, the thin grey curves are the temperature profiles of just a few hundred stacks of 50 random positions. The blue solid curve is the result from the fiducial stacking of Gr08 voids. Every stacking here is carried out in the WMAP V CMB map, smoothed at 30.6'. \textit{Bottom panel:} Same legend as before for the photometry profiles. The results from the V (solid blue) and W (dashed blue) band are also shown. Similarly to Gr08, the signal from their voids stands out at more than $3\,\sigma$ on an angular scale of $4^\circ$.}
\label{fig:SNR1}
\end{figure}

\begin{figure}[t]
\centering
\includegraphics[width=0.45\textwidth]{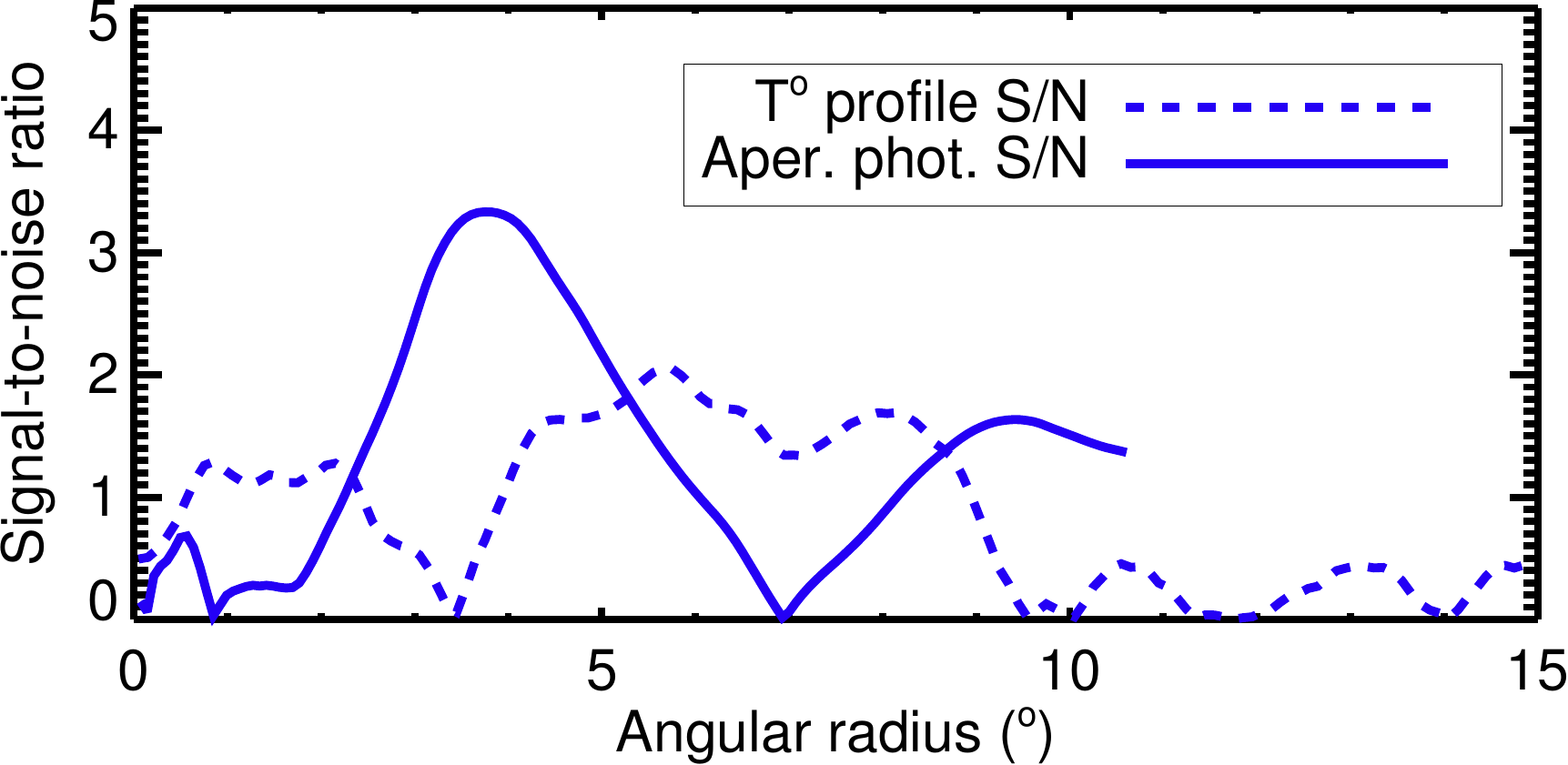} \\
\includegraphics[width=0.45\textwidth]{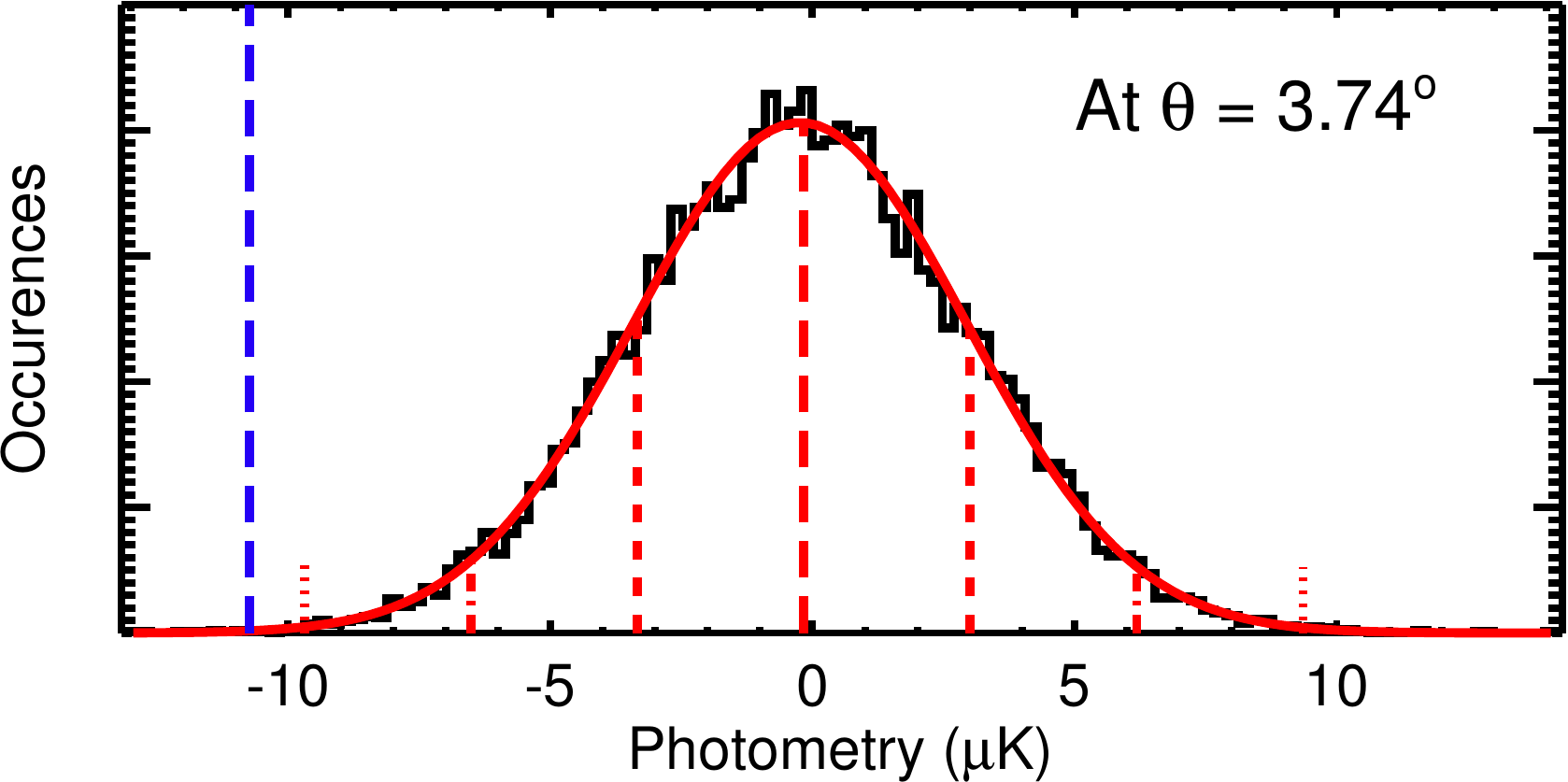}
\caption{\small\textit{Top panel:} Signal-to-noise ratio (as defined in Eq.~\ref{eq:SNR}) of the temperature (dashed blue) and photometry (solid blue) profiles of our fiducial stack in WMAP V CMB map. A $\sim3.3\,\sigma$ signal stands out in the photometry at a scale of $3.74^\circ$. \textit{Bottom panel:} Distribution of photometry values for an aperture angle of $3.74^\circ$, from 14000 stacks of 50 random positions. The mean, $1$, $2$, and $3\,\sigma$ values are marked the same way as in Fig.~\ref{fig:SNR1}. The blue long-dashed line shows the value obtained from the fiducial stacking. The best fitting Gaussian (red solid curved) follows closely the distribution.}
\label{fig:SNR2}
\end{figure}

An application of this procedure is illustrated in Figs.~\ref{fig:SNR1}~and~\ref{fig:SNR2} where we assess the significance of our fiducial stack (Gr08 voids) in the CMB of WMAP V band. For this particular example, we used more than 14000 sets of 50 random positions, which seems enough to sample the distribution of temperature and photometry profiles. Indeed this is hinted at by Fig.~\ref{fig:SNR2} where the histogram of photometry values at a given angular size follows a Gaussian distribution closely. We discuss the interpretation of this data in Sects.~\ref{sec:4} and \ref{sec:5}. The procedure for this estimation of the significance is robust and is used in the next section for all our results. 
17193

\section{Results}
\label{sec:4}
\begin{figure}[t]
	\centering
	\includegraphics[width=0.49\textwidth]{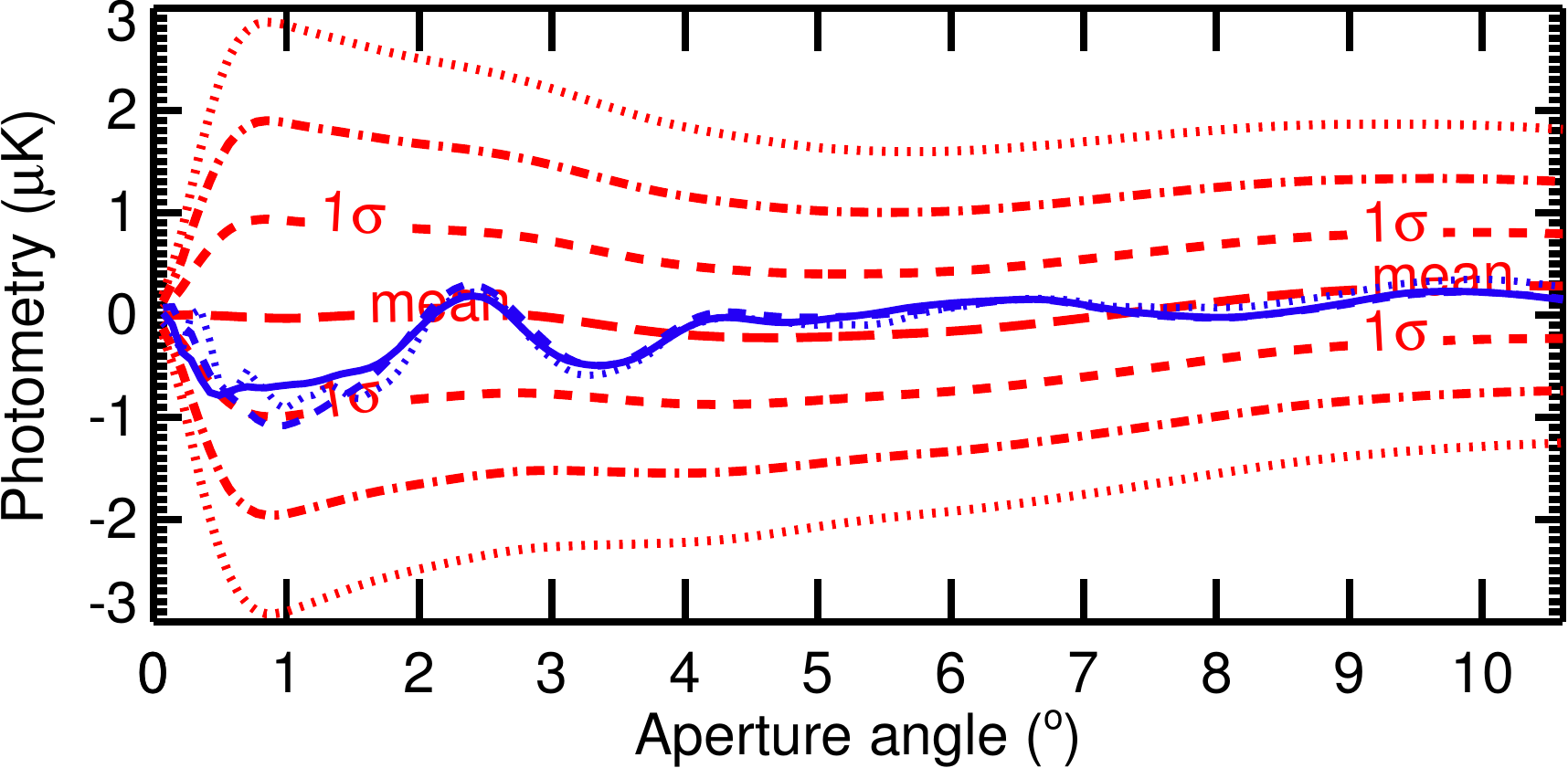}
	\caption{\small Photometry profiles from the stacking of Pan et al.\ voids (same conventions as Fig~\ref{fig:SNR1}).}
	\label{res:Pan}
\end{figure}

\begin{figure*}[!ht]
	\centering
	\includegraphics[width=0.98\textwidth]{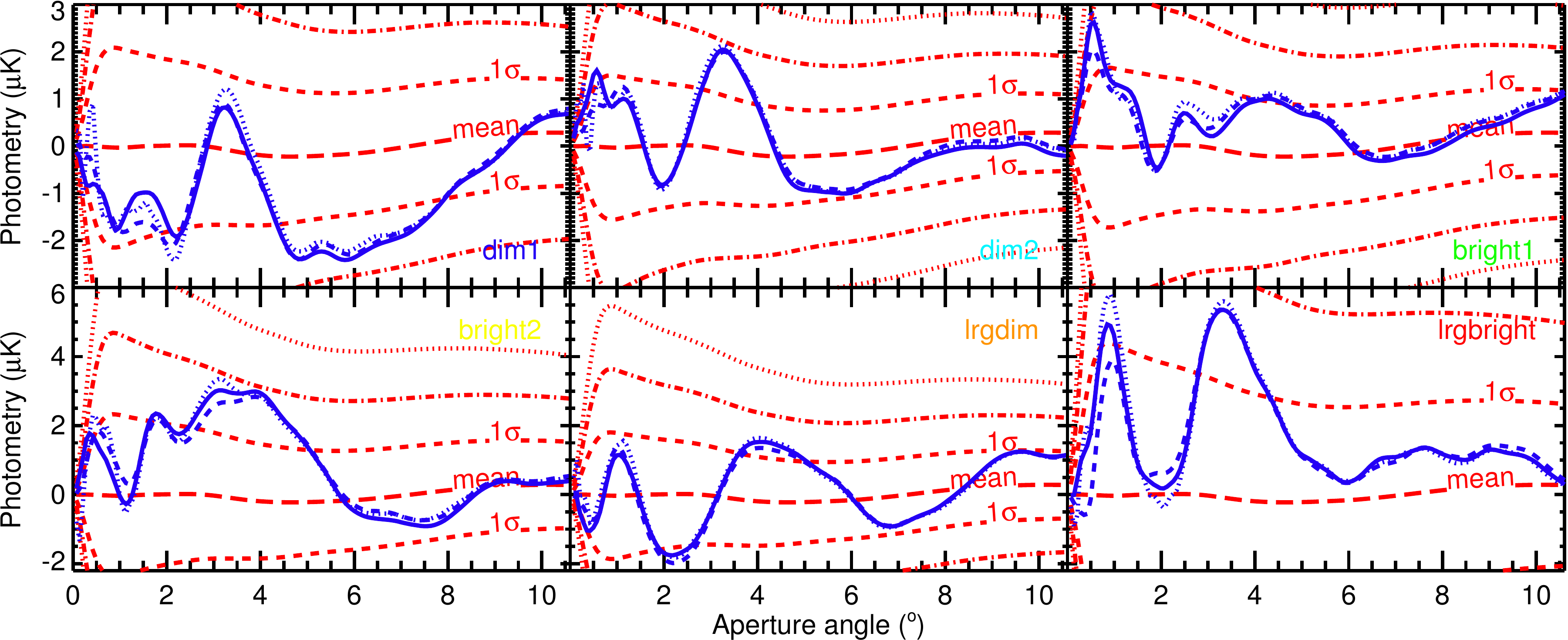}
	\caption{\small Photometry profiles from the stacking of Sutter et al.\ voids and its six subsamples (same conventions as Fig~\ref{fig:SNR1}).}
	\label{res:Sut}
\end{figure*}

We apply the procedure described in the previous section to all our stacks in order to estimate their significance. We show the results for the photometry of the stacks for each catalogue and subsample, with an assessment of the significance (Figs.~\ref{fig:SNR1}, \ref{res:Pan}, and \ref{res:Sut}). 

\subsection{Revisiting Gr08}
\label{sec:4-1}

The work of Gr08 reported a $3.7\,\sigma$ signal ($-11.3\,\mu$K) in the stacking of their voids on a scale of $4^\circ$. With the same dataset we find a reasonable agreement (Fig.~\ref{fig:SNR2}) with a preferred scale of $\sim3.7^\circ$ with an $\mathrm{S/N}\sim3.3$ ($-10.8\,\mu$K). These differences can be imputed to our use of WMAP 7 maps instead of the WMAP 5 ILC map for Gr08 and, to a lesser extent, to light differences in the stacking procedure, profile calculations, or significance estimation. While we can argue about its cosmological origin, the signal seems to be persisting and is essentially identical across frequencies (see Fig.~\ref{fig:SNR1}) as expected for the iSW effect. However, we found an important feature in the temperature profile of the stacked image and its significance. Indeed, in the top panel of Fig.~\ref{fig:SNR1} we see that the central cold spot of the signal (below $3.5^\circ$) does not particularly stand out compared to random stacks ($1\,\sigma$ significance only). On the other hand, we measure  a wide hot ring with around the spot a higher significance (up to $2\,\sigma$) on scales between $3.5^\circ$ and $10^\circ$, clearly visible in the middle image of Fig.~\ref{fig:poles}. The impact of this ring is even visible in the photometry profile on higher scales, with a significance reaching almost $2\,\sigma$ around $9^\circ$.

Interpreting it in the light of the iSW effect, this would imply the presence of much higher overdensities surrounding the already large supervoids. Considering the filamentary structure of our Universe, this situation is unlikely, and the source of this hot ring remains unknown. This peculiarity leads us to question whether the measured central cold spot -- physically interpreted as an iSW signal -- is really remarkable. It might as well be due to random fluctuations of the CMB, of which the significance in the photometry profile is coincidentally strengthened by a surrounding hot region in the stacked image. 

One could find an argument in favour of the iSW interpretation by noting the good match between the preferred scale in the S/N ($\sim3.7^\circ$) and the mean effective angular radius of the Gr08 voids ($\sim3.5^\circ$). As a matter of fact, the same argument can be turned around to rebut this interpretation. We intuit that the iSW effect should fade close to the border of the voids, making the cold spot noticeably smaller than the underlying structure. However, the presence of the wide hot ring around the central cold spot artificially pulls the preferred scale towards higher values, making it difficult to interpret the apparent match of scales. In a similar line of thought, note that the same analysis performed on the Gr08 \emph{superclusters} gives a photometry profile that peaks on angular scales more than twice as large as those of the underlying clusters \citep{Planck_ISW}. 

Before any definitive answer on this matter can be reached, this discussion requires a complementary rigorous investigation, through theory and/or numerical simulations, of the iSW effect expected from such superstructures, which is beyond the scope of this paper.

\subsection{Other catalogues}
\label{sec:4-2}

The two other catalogues used in our analysis yield mixed results, with no significance on the same level as the Gr08 catalogue. While discouraging, one has to remember that according to current theoretical predictions, we expect this iSW signal to be difficult to detect.

The stacking of the 1055 voids of Pan et al.\ gives a faint signal below the $1\,\sigma$ level at about $1^\circ$ (Fig.~\ref{res:Pan}), which does not allow any interpretation. This may be due to the high number of voids in the sample, which has a very wide distribution of angular sizes. We considered the idea of dividing this catalogue into subsamples based on redshift, radius, and/or angular sizes. After several attempts, it did not yield any significant result, and we faced a number of issues that include the hassle of finding appropriate bins of size for the subsamples and possible a posteriori selection effects. Indeed, a smaller number of structures implies a narrower range of sizes and redshifts, but it can also greatly reduce the S/N. 

We still addressed this issue partially by  separately analysing the six subsamples of Sutter et al.\ (Fig.~\ref{res:Sut}) -- a simple division based on redshift but in principle free of selection effects. From this analysis, only one of the subsamples (\textit{dim1}) yields a negative signal in the photometry with a significance higher than $1\,\sigma$. The other profiles are either entirely positive (\textit{lrgbright}), or their significance is below $1\,\sigma$. An explanation for this apparent absence of significant results may be found by considering the dispersion in the angular radii of the voids. The subsamples from Sutter et al.\ contain significantly more objects compared to the Gr08 sample, but their sizes on the sky are much more scattered (see Fig.~\ref{fig:voids:all}). Mixing such a variety of void sizes necessarily dilutes the associated iSW signal over the same range of scales, drastically reducing its significance.

This preliminary analysis strongly suggest that we need to take the size of each individual void  into account in the stacking procedure to improve the significance of the results.

\subsection{Rescaling tests}
\label{sec:4-3}

In light of the results above, we adopted another approach in order to enhance the significance of the signal. We kept each subsample in its entirety and did  the stacking analysis again, but this time rescaling the voids according to their effective radii. In practical terms, this means cutting the CMB patches so that each void occupies the same space on the stacked image.

We keep the same protocol as described in Sect.~\ref{sec:3-1}, but we change the resolution of the extracted patches, so that for each patch, it now depends on the size of the corresponding void. Each time, we cut a square patch with a side six times the size of the void effective radius it contains. Naturally, we also adapt our protocol for the estimation of the significance: The sets of random positions are still drawn the same way, but each random patch is first subjected to the same rescaling as done on the identified void patches. This will have the effect of mixing different scales in the original CMB map, and will most likely amplify the variance of these random stacks compared to those of Sect.~\ref{sec:3-3}.

\begin{figure}[t]
	\centering
	\includegraphics[height=3.8cm]{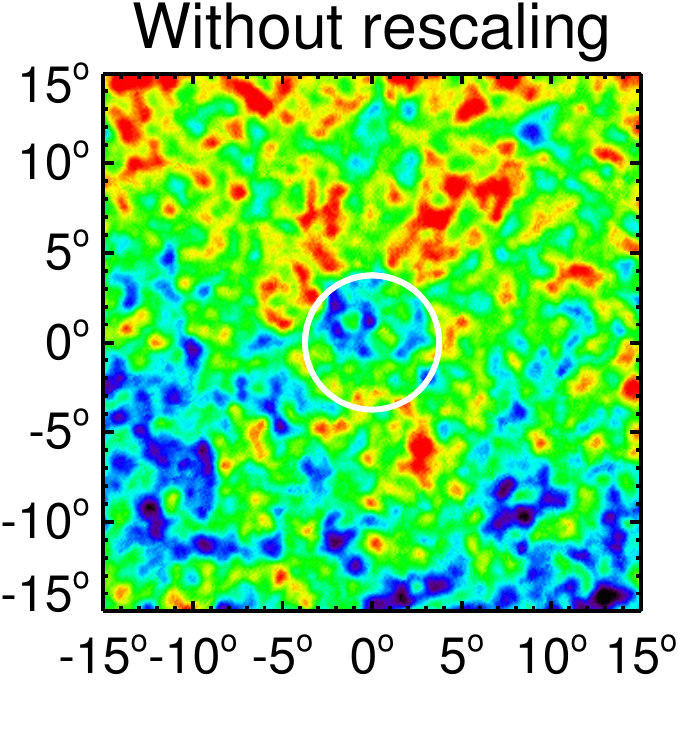} \hspace{0.3cm} \includegraphics[height=3.8cm]{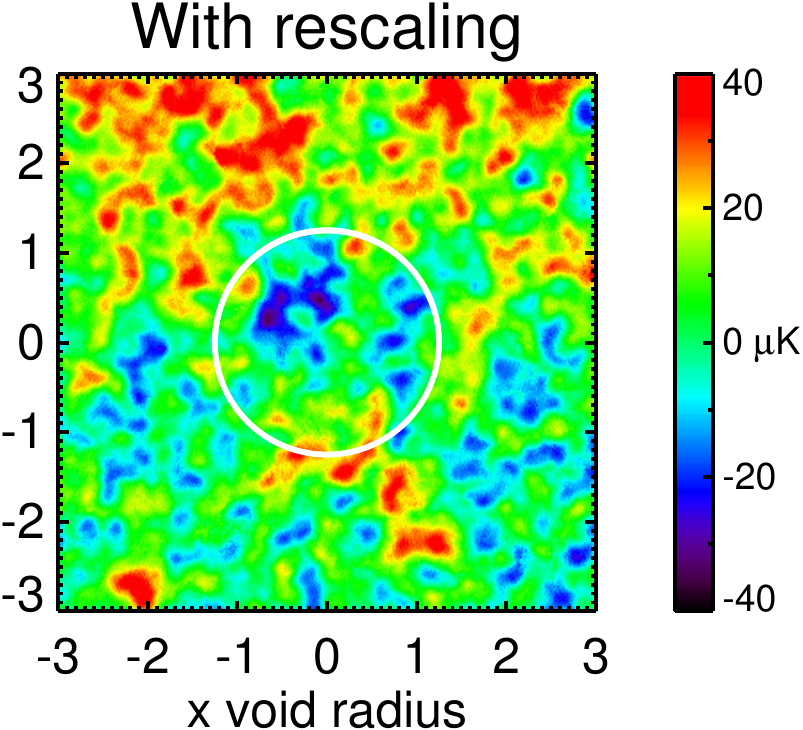} 
	\caption{\small Stacked images of Gr08 voids, without (\textit{left}) and with (\textit{right}) a rescaling of the CMB patches proportional to the void sizes.}
	\label{fig:rsc_image}
\end{figure}

\begin{figure}[t]
	\centering
	\includegraphics[width=0.49\textwidth]{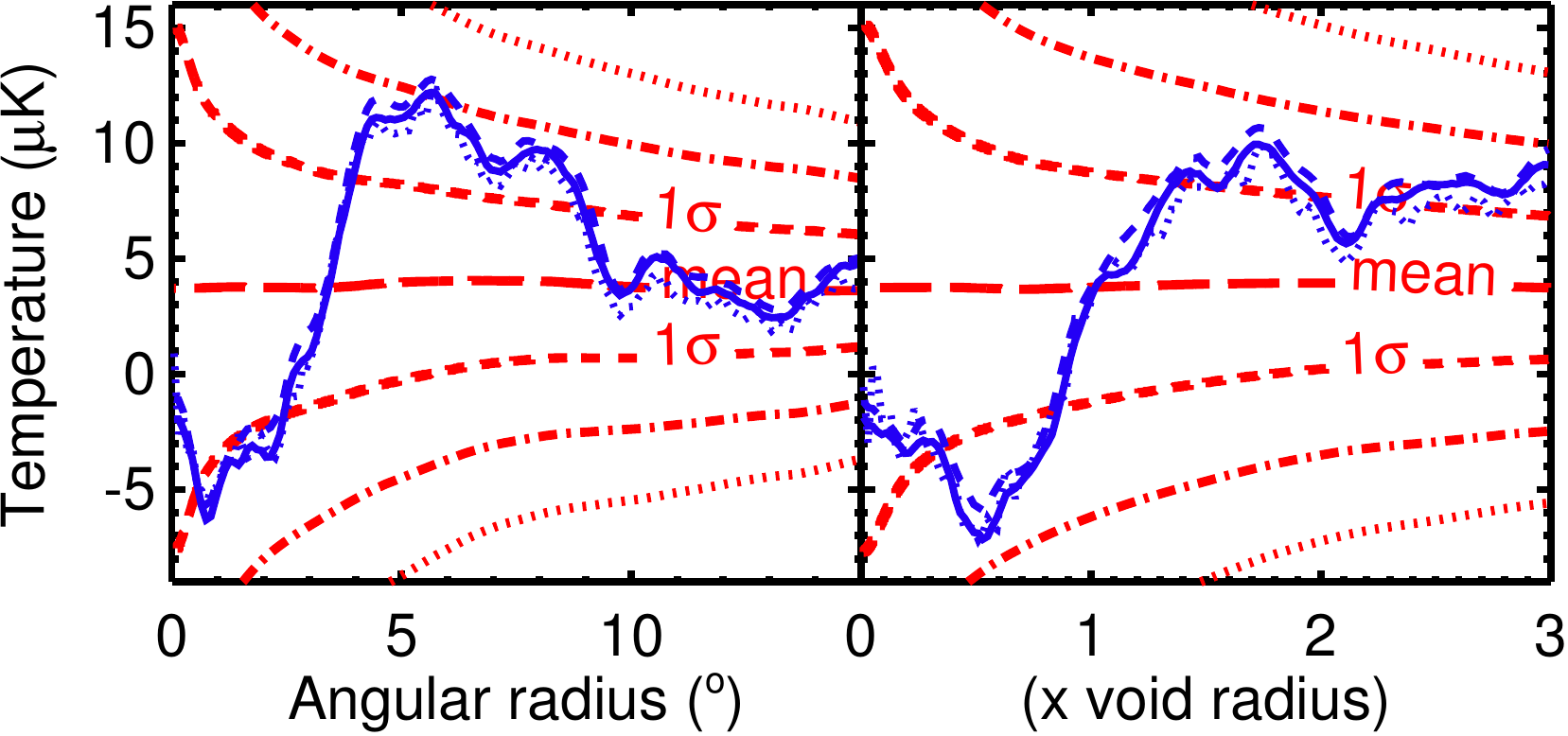} \\ \ \\
	\includegraphics[width=0.49\textwidth]{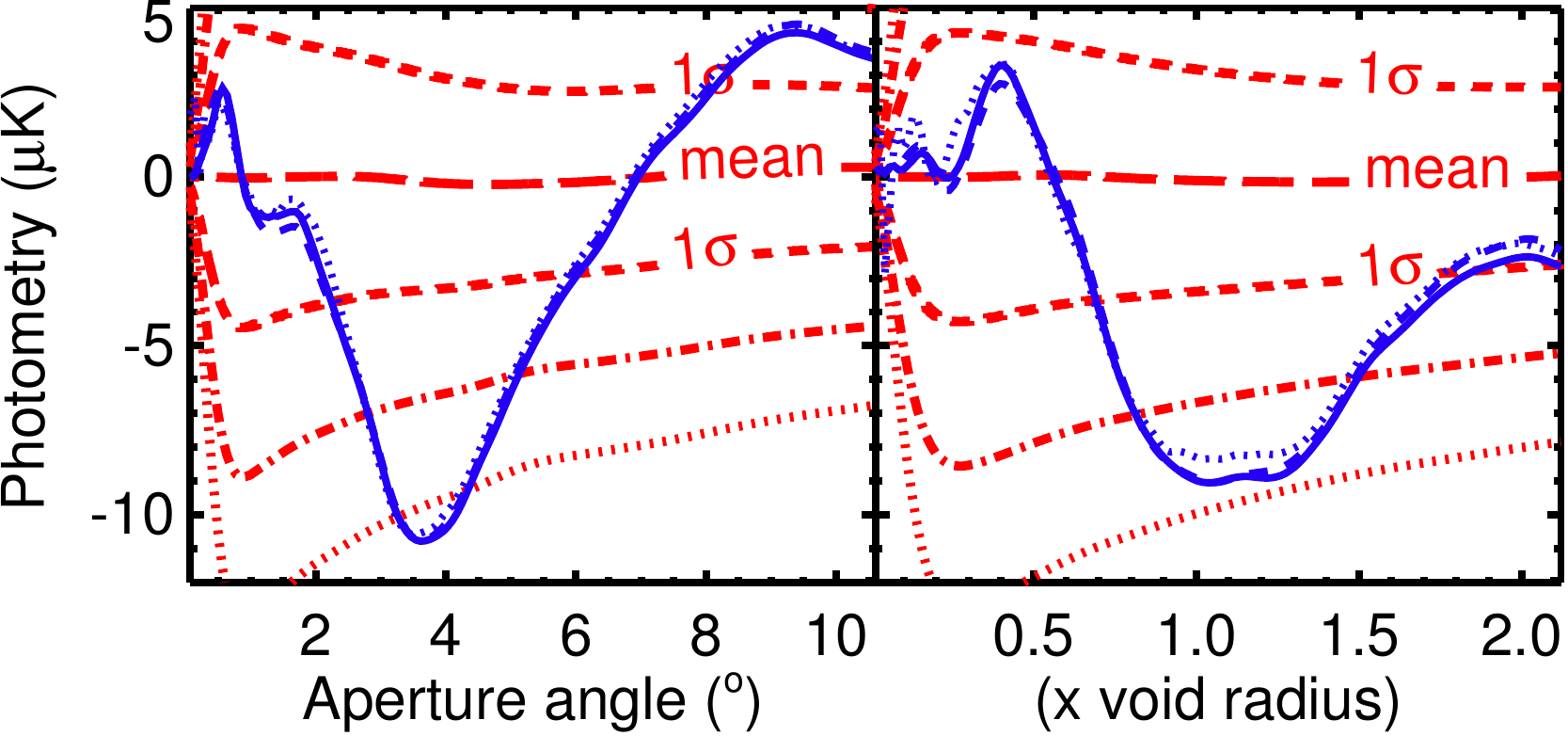}
	\caption{\small Temperature (\textit{top}) and photometry (\textit{bottom}) profiles for the original (\textit{left}) and rescaled (\textit{right}) stacking of Gr08 voids (same conventions as Fig~\ref{fig:SNR1}).}
	\label{res:Gra_rsc}
\end{figure}

We begin with the fiducial stacking of the voids of Gr08. A comparison between the stacked images is illustrated in Fig.~\ref{fig:rsc_image} while the different profiles are shown in Fig.~\ref{res:Gra_rsc}. The signal identified in Sect. \ref{sec:4-1} still appears after rescaling, with the best significance around scales between 1 and 1.3 times the void effective radii. This value somehow seems a little too high since, as stated above, we expect this value to be around or smaller than one (due to the irregular geometry of the voids). The significance of the signal is also found to be lower (S/N$\sim2.8$ versus $\sim3.3$). This is partly due to the increased variance of the signal (cf. the wider $1/2/3\,\sigma$ limits in the profiles) induced by the rescaling. But it is also a consequence of the lower amplitude measured for the signal, at odds with our expectations of the rescaling procedure. This could be a further hint that random CMB fluctuations actually contribute notably to the signal seen in the stacked image. On the other hand, the temperature profile of the rescaled stack is closer to expectations, with a higher significance cold spot in the centre (cf. first paragraph of Sect.~\ref{sec:4-1} for comparison).


\begin{figure}[t]
	\centering
	\includegraphics[width=0.49\textwidth]{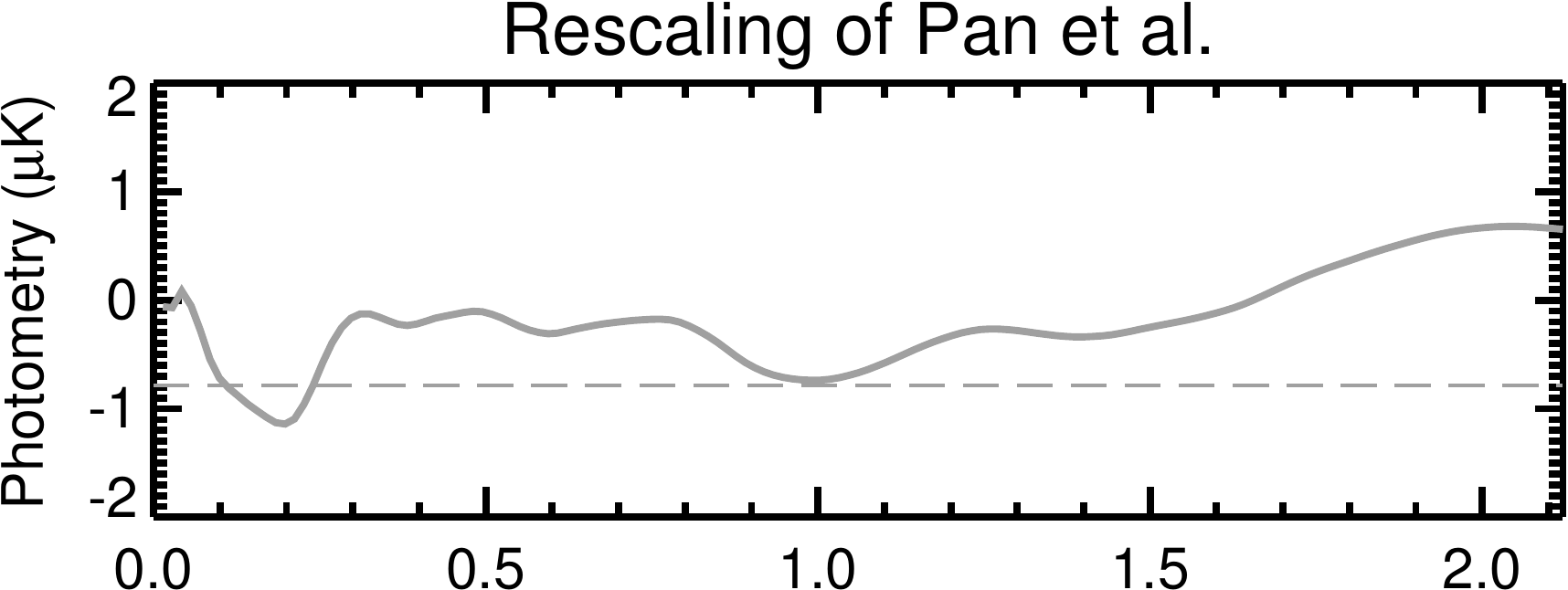} \\ \vspace{0.3cm}
	\includegraphics[width=0.49\textwidth]{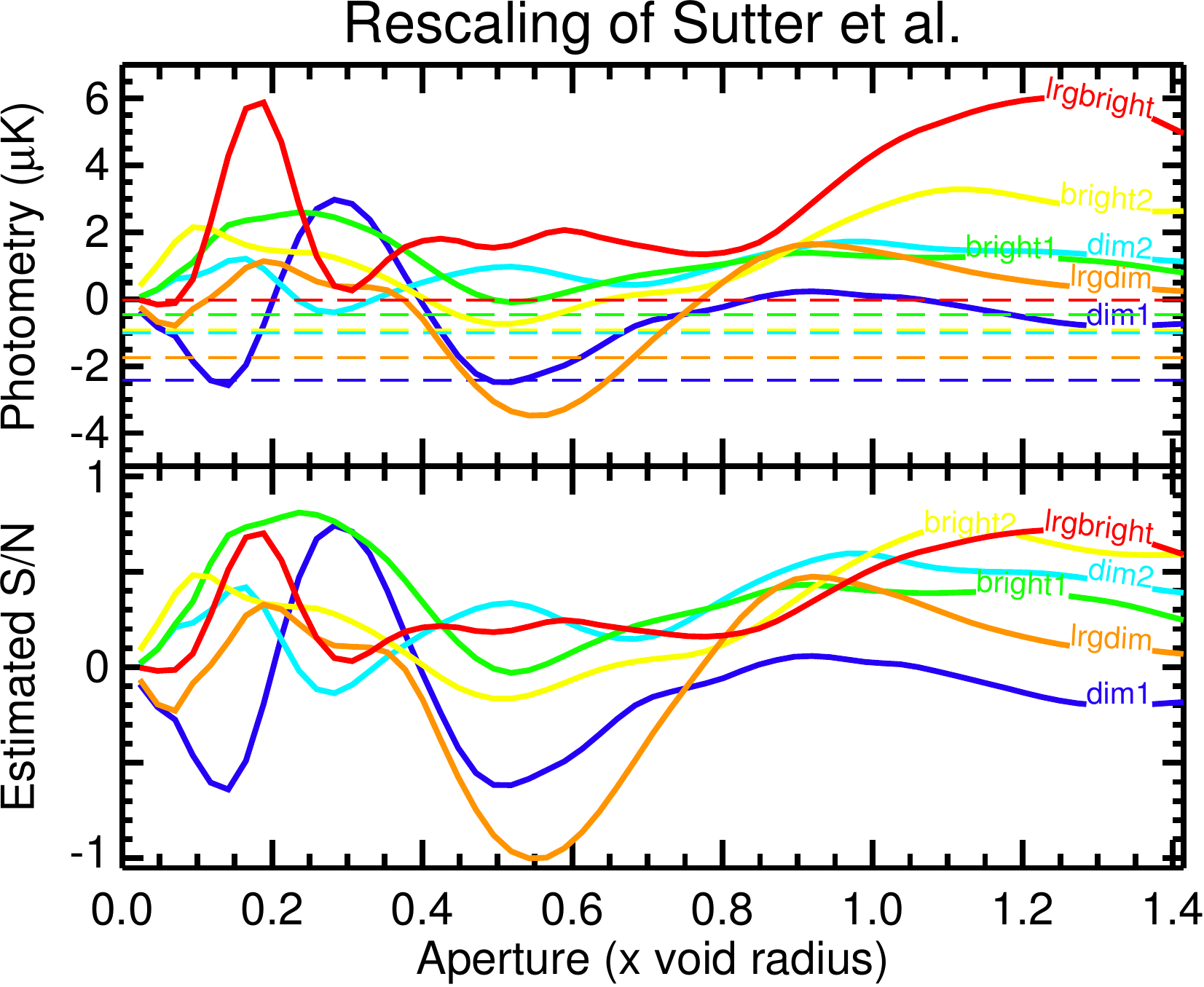}
	\caption{\small Summary of the photometry profiles extracted from the rescaled stacks of Pan et al.\ (top) and Sutter et al.\ (middle), performed in the WMAP V band cleaned map. The coloured dashed lines indicate for each sample the lowest amplitude measured in the original stacked image (without rescaling the voids). These allow to roughly estimate if the rescaling procedure did improve the detection of any previously detected signal. The bottom panel shows the previous profiles multiplied by $\sqrt{N_v}$, with $N_v$ the respective number of voids in each subsample. They are then normalized to the strongest signal (\textit{lrgdim}). These curves provide an estimate of their potential significance as the noise in the stacked image is expected to scale as $1/\sqrt{N_v}$ approximately.}
	\label{res:Sut_Pan_rsc}
\end{figure}

From this first test, we understand that if the rescaling process does not at least improve the absolute amplitude of any previously detected signal, then any subsequent significance estimation is very unlikely to yield an improvement since the necessarily larger variance decreases the S/N further. Therefore we first produced an overview of the photometry profiles obtained from the rescaled stacks of Sutter et al.\ and Pan et al.\ voids (top of Fig.~\ref{res:Sut_Pan_rsc}). Again, no signal of particular importance arises from this new analysis of Pan et al.\ voids (except at very small angular sizes, most likely due to random fluctuations and not in relation to any underlying structure). Concerning the Sutter et al.\ catalogue, signals seem to arise in several of the rescaled profiles, especially on a scale equal to 0.5-0.55 times the voids effective radii, with a clear departure from the previous (without rescaling) results for some of them, such as the \textit{lrgdim} subsample. However, some of the other subsamples (\textit{bright1}, \textit{lrgbright}) do not benefit from the rescaling procedure. 

\begin{figure}[t]
	\centering
	\includegraphics[width=0.49\textwidth]{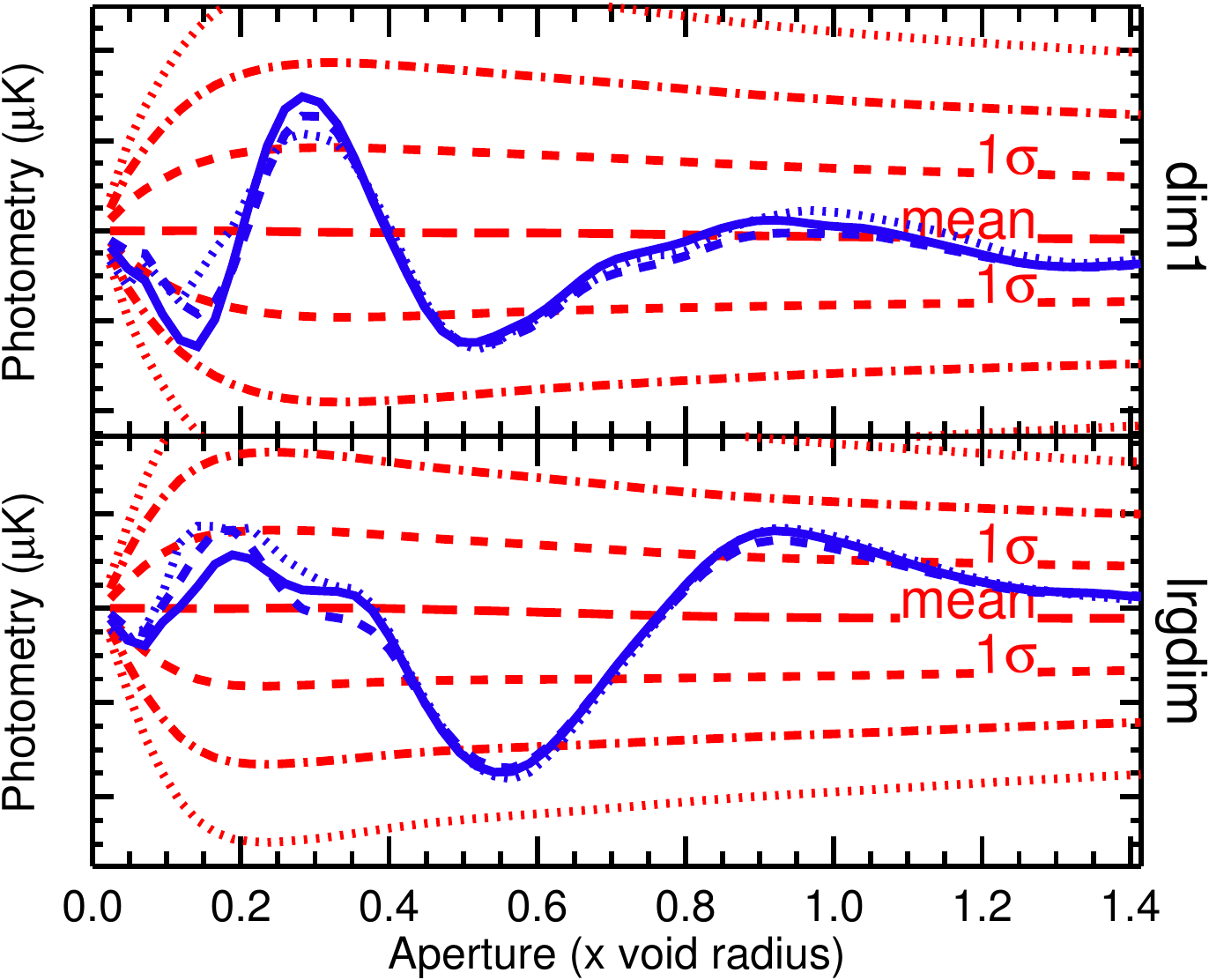}
	\caption{\small Photometry profiles and significance for the rescaled stacks of two Sutter et al.\ subsamples, \textit{dim1} and \textit{lrgdim} (same conventions as Fig~\ref{fig:SNR1}).}
	\label{res:Sut_2_rsc}
\end{figure}

We identified the two most promising subsamples,  \textit{dim1} and \textit{lrgdim}, and we evaluated the significance of their rescaled profiles. The results, shown in Fig.~\ref{res:Sut_2_rsc}, are close to our expectations. The \textit{dim1} subsample yields a signal at $\sim 0.52$ times the void effective radius, with a significance similar (albeit a bit smaller) to the  results without rescaling (around $1.36\,\sigma$). This is coherent with the fact that the amplitude of the signal remained almost at the same level (illustrated by the corresponding dashed line in Fig.~\ref{res:Sut_Pan_rsc}), whereas the rescaling procedure slightly increased the variance of the random stacks used in the S/N estimation. We note that the apparently significant signals at small aperture angles ($\sim 0.15$x and 0.3x) are not constant across frequencies and therefore are probably not related to an iSW effect. Regarding the \textit{lrgdim} subsample, it gives a $\sim2.35\,\sigma$ signal around 0.57 times the void radius, a clear improvement over the non-rescaled results that was expected considering the stronger amplitude of the rescaled signal. For both subsamples, we note that the amplitude of the highlighted signal stays remarkably constant across frequencies, whereas the other potentially significant features of the profiles (on smaller scales notably) are not achromatic. 

Interestingly, we observe that the analysis of both subsamples presents an iSW-like signal at a scale slightly above half the effective radius of the considered voids. Although this clearly differs from the results obtained with the Gr08 catalogue, a physical explanation can be found in the geometry of the voids from these subsamples. Indeed, as noted by \cite{Sutter2012}, the majority of them present a similar shape to a prolate ellipsoid with an ellipticity close to two. Since the orientation of these voids is a priori random, we can intuit that stacking such ellipsoids (and their associated iSW signature) will eventually give a circular signal, on a smaller typical scale, closer to half the major axis of the ellipsoids. An accurate estimation of this scale would require an extensive analysis of the individual geometry of the considered voids, and their associated iSW signature. Nonetheless, a recent study by \cite{Cai2013}\footnote{We note that Cai et al.\ have withdrawn their paper since their observational results were based on an older version (August 2012) of the Sutter et al.\ void catalogue which was plagued by a few issues -- see \url{http://www.cosmicvoids.net}. The theoretical conclusions of Cai et al.\ based on their numerical simulations remain however valid.} comforts our intuition. They constructed large N-body simulations following a $\Lambda$CDM cosmology and identified the voids in the manner of \cite{Sutter2012} and Gr08. They subsequently computed the iSW signature resulting from the stacking of their simulated voids, which highlighted an optimal scale around 0.6 times the effective radius of the voids for the analysis by aperture photometry. The similarity between this predicted scale and the signal we observe in the Sutter et al.\ void subsamples is encouraging. In contrast, we should note that using a different approach, \citet{Carlos2012} generated matter density maps (from both Gaussian and N-body simulations) that followed the redshift distribution of Gr08, from which they derived full-sky iSW maps. After performing a stacking in these maps at the locations of the density peaks/troughs, they found that the scales highlighted by aperture photometry ranged from $1^\circ$ to $20^\circ$ -- with a maximum around $7^\circ$, but a small amplitude of $\sim 2\,\mu$K.

\subsection{Alternative approach with rescaling}
\label{sec:4-4}

From the previous sections, we gathered that in some of the Sutter et al.\ subsamples, an iSW-like signal seems to appear around 0.5-0.6 times the voids effective radii and that it is especially significant in the \textit{lrgdim} sample. A possible explanation may come from the presence in this particular subsample of some of the largest voids in the whole Sutter et al.\ catalogue (as can be seen in Fig.~\ref{fig:voids:all}), which are supposed to yield the strongest iSW effect. The \textit{lrgbright} sample also contains many large voids, but because of its small number of objects (50), the level of noise from primordial CMB fluctuations is still high, hence the absence of a significant signal in the stacked image. 

\begin{figure}[t]
	\centering
	\includegraphics[width=0.49\textwidth]{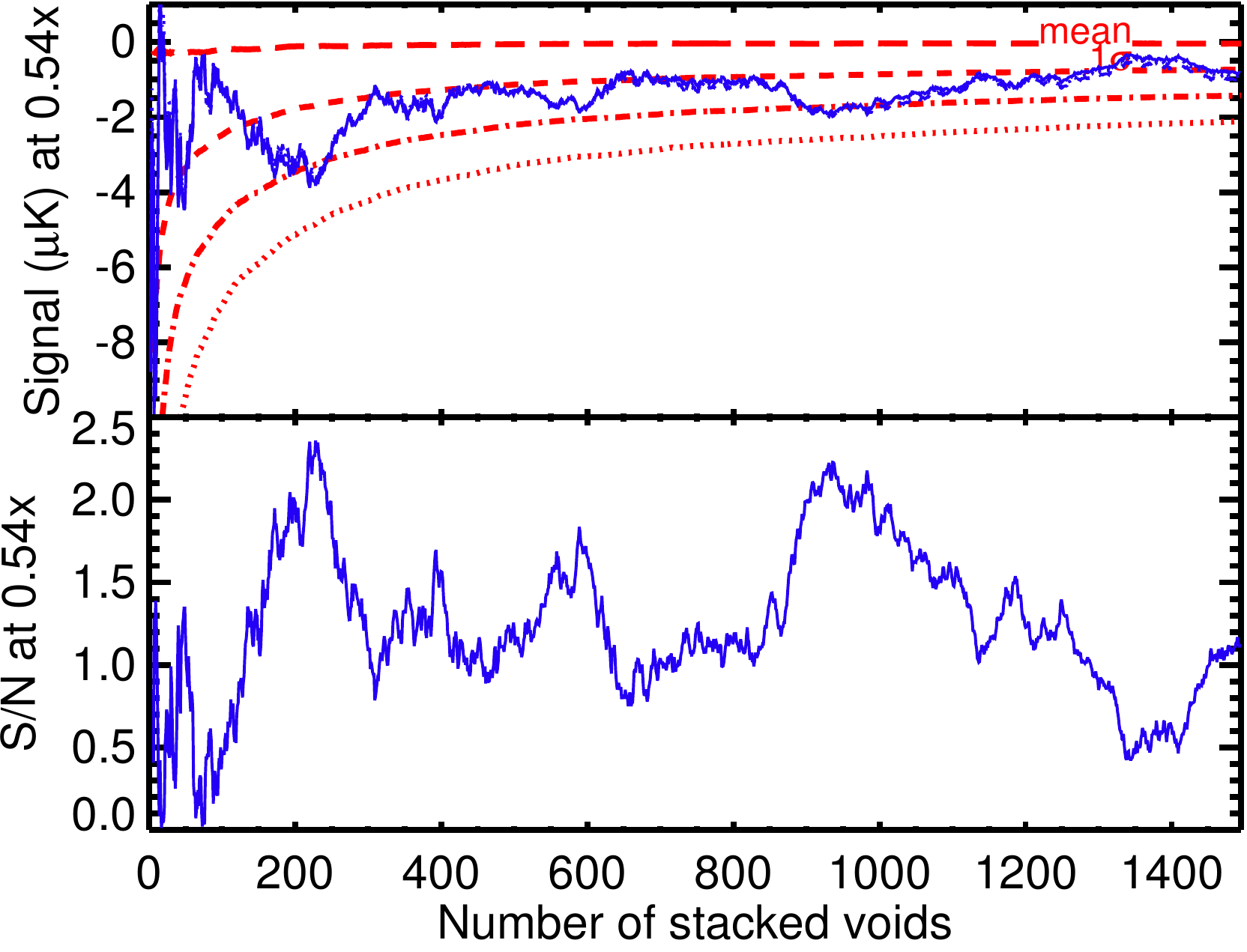}
	\caption{\small \textit{Top:} Photometry values of the stacked and rescaled voids from the Sutter et al.\ catalogue, at an aperture of 0.54 times their effective radii, as a function of the number of voids (sorted by decreasing size). Significance contours are computed using random stacks produced from the WMAP V band cleaned map. Legend is identical to previous plots of photometry profiles (see Fig~\ref{fig:SNR1}). \textit{Bottom:} For the stacking performed in the WMAP V map, S/N of the above photometry values computed using the significance contours.}
	\label{res:SNR_Sut_rsc}
\end{figure}

This indicates that instead of considering each subsample separately, a better approach may be to combine them all and stack the voids starting from the largest ones. Indeed, in theory the noise should scale as usual roughly as the inverse square root of the number of stacked voids, but the stacked iSW signal is also expected to drop at some point due to the addition of smaller and less contributing voids. By starting from the largest voids, we intend to select the supposedly largest iSW contributions in order to keep the stacked signal from dropping too fast and effectively to boost the S/N of the detection. We carried out this analysis on the 1495 voids of Sutter et al., first focusing on the whole photometry profiles and increasing progressively the number of stacked voids. As expected, a negative signal consistently appears around an aperture of 0.54 times the voids effective radii. As intuited before, its amplitude gradually decreases as we include smaller and smaller voids in the stacking. To estimate the significance of this signal, we focus on the value of the photometry at this particular aperture scale. In the top half of Fig.~\ref{res:SNR_Sut_rsc}, we show these values as a function of the increasing number of stacked voids. Similarly to the previous section, we estimate the significance of these values by repeating the analysis many times after randomly shifting the stacked positions. Therefore we can compute the S/N of these results, shown in the bottom half of Fig.~\ref{res:SNR_Sut_rsc}. We note once again that the photometry is stable across frequencies and consistently negative for practically any number of stacked voids, but the shape of this curve and its significance are hard to interpret. The significance first rises up to $\sim 2.3\,\sigma$ for the first 200 stacked voids, a behaviour that would be expected from an ISW signal that progressively takes over the CMB noise. After this, the S/N quickly decreases and then oscillates between about $1\,\sigma$ and $2\,\sigma$ before dropping, after stacking more than 1300 voids. Although this significance appears to vary quite significantly, the stability of the signal itself (always negative and on the same scale) may indicate that this variability is due to random CMB fluctuations. 

\begin{figure}[t]
	\centering
	\includegraphics[width=0.49\textwidth]{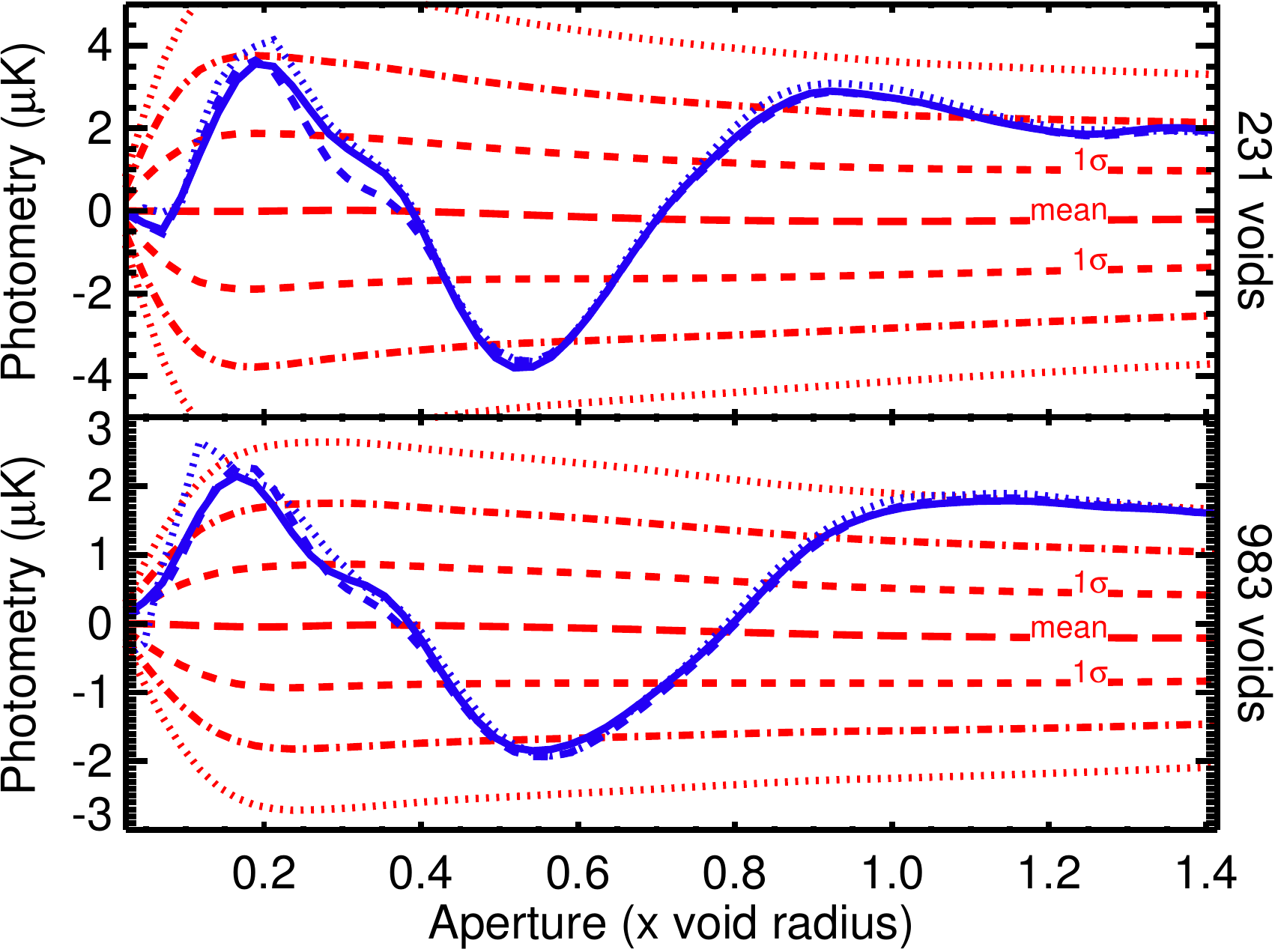}
	\caption{\small Photometry profiles and significance of the rescaled stacking of the 231 (top) and 983 (bottom) largest voids from the Sutter et al.\ catalogue (same conventions as Fig~\ref{fig:SNR1}).}
	\label{res:sig_nSut}
\end{figure}

We selected two particular numbers of voids with a high significance observed in Fig.~\ref{res:SNR_Sut_rsc}, namely 231 and 983 voids, and performed the same photometry/significance analysis as in Sect.~\ref{sec:4-3}. The results are shown in Fig.~\ref{res:sig_nSut}: both profiles highlight a scale again equal to 0.54 times the voids effective radii where the signal reaches significances equal to $2.38\,\sigma$ for the 231 voids and $2.20\,\sigma$  for the 983 voids. Despite having more voids, the small drop in significance may again be caused by CMB fluctuations and/or the inclusion of small voids. Another notable feature appears in both photometry profiles, namely one hot significant signal on smaller scales ($\sim 0.2$x) and one on larger scales ($>0.9$x). The former and its associated scale seem to coincide with an angular size close to $1^\circ$. It is directly related to the curious fact that the minimal temperature spots do not appear at the centre of the stacked image. Were the stacks centred on the coldest spot, we would not have obtained this positive signal in the photometry profiles at small angles. 

The origin of this hot central spot is undetermined, but could be due to several contributions of which the background CMB fluctuations, the irregular shape of the underlying voids, and the possible mismatch between the position of the void barycentres and their most underdense zones. Concerning the signal on larger scales, what comes to mind is the possible influence of large scale fluctuations through the low multipoles of the CMB, already glimpsed in \ref{sec:3-2-3}. Thus we redid the stacking of the same sets of voids on new maps with a few low multipoles removed from $\ell=2$ to 20.  The results then showed that these multipoles indeed make a non-negligible contribution to our photometry profiles: removing them noticeably reduces the measured amplitude on large scales but keeps the rest of the signal almost intact. Although it does not account for the entirety of the large scale signal, it does reduce its significance to less remarkable levels.

In summary, the introduction of a rescaling in the stacking procedure yielded different results depending on the considered samples of voids. Concerning the Gr08 catalogue, the rescaling gave a more coherent shape to the temperature profile of the stacked image: a significant cold spot and a less significant hot ring. However, it highlighted a tension between the scale preferred by the aperture photometry and the actual smaller size of the stacked objects, and also slightly (but counterintuitively) reduced the significance of the detected signal. These may be hints that this signal is not entirely an iSW effect produced by voids, but is partly due to random CMB fluctuations. This would explain both the scale discrepancy (CMB fluctuations and voids are uncorrelated) and the lower S/N (the noise from the CMB is ``distributed" across all scales by the rescaling).

On the other hand, the rescaling process had positive results on the much larger catalogue of Sutter et al., highlighting a particular scale around half the void sizes in all the tests performed, in apparent agreement with both intuitive arguments and theoretical works in the literature. Although the maximum observed significance only reaches around $2.3\,\sigma$ and the signal depends quite significantly on the number of stacked voids and their size, the persistent nature of the signal seems to bolster the case for iSW detection.

Lastly, we note that the Pan et al.\ catalogue did not benefit from the rescaling process, and wish to mention that we also explored the same approach used with the Sutter et al.\ catalogue (gradual stacking starting with the largest voids), but it did not yield any significant results. We hypothesise that this is caused by the narrow range of small sizes of these voids, whose faint iSW signal is likely to be dominated by CMB fluctuations.
 
23553

\section{Discussion}
\label{sec:5}
\subsection{Selection effects}
\label{sec:5-1}

To further elaborate on the possible conclusions that we can deduce from the present work, there are a couple of effects that can make the interpretation of the results even more difficult and that we should keep in mind. We now mention a few of them.

\begin{figure}[t]
	\centering
	\includegraphics[width=0.49\textwidth]{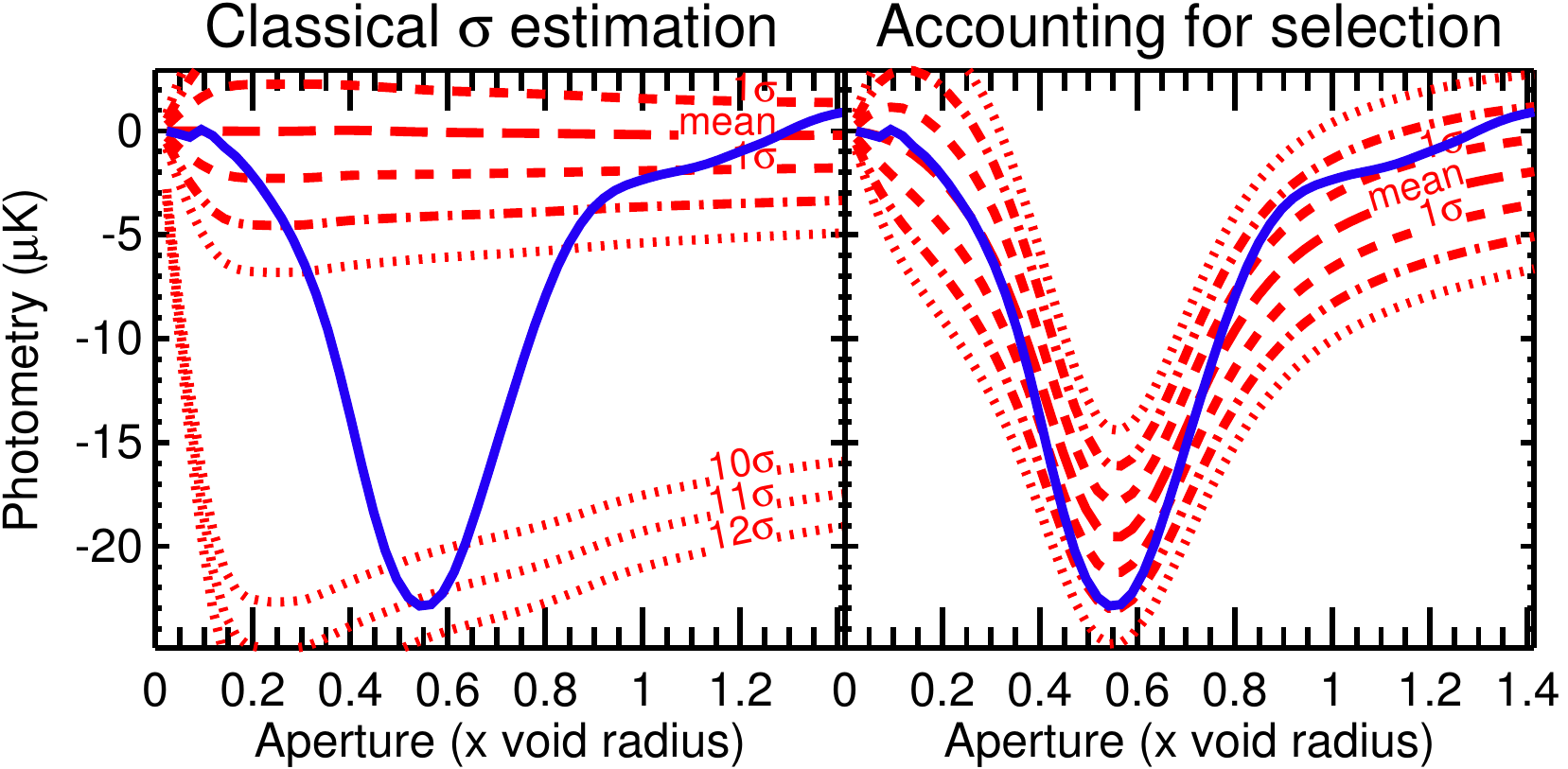}
	\caption{\small Photometry profile for the rescaled stacking of half the Sutter et al.\ \textit{lrgdim} subsample (146 voids out of 291), chosen so that the amplitude at 0.56 times the voids radii is the strongest (see text for details). \textit{Left:} the significance is estimated with stacks of 146 random positions. \textit{Right:} the significance accounts for the selection effect. The difference between the two significances is obvious and very pronounced.}
	\label{res:selec}
\end{figure}

The results from the previous section highlight the difficulty of getting a clear detection of a signal, at least when using the various catalogues as such. With this in mind, one could be tempted to amplify the signals hinted at by isolating the voids that contribute most. We experiment with this idea and apply it to the best S/N from our new results in Sect.~\ref{sec:4-3}, i.e.\ the rescaled stacking of the \textit{lrgdim} subsample. From the 291 voids of the original set, we keep only the half (146) that contributes the most to the minimum of the photometry profile on the scale of 0.56 times the voids radii. The resulting photometry profile is plotted in Fig.~\ref{res:selec} and indeed shows a dramatic increase in the amplitude of signal. From there, the crux of the matter is to assess the significance of this new result. When using the same procedure as in Sect.~\ref{sec:3-3} (many stacks of 146 random positions), we obtain a surprisingly high S/N, above $11\,\sigma$. This value clearly overestimates the real significance, since it ignores the selection that we performed on the sample. We need to revise our protocol as follows. We first generate many sets of 291 random positions. Then for each set, we select and stack only the half that contributes most to the photometry profile on the same scale of 0.56 times the voids radii. Once we draw enough such random stacks, we keep the rest of the procedure identical. The corrected significance (Fig.~\ref{res:selec}) drops down to a level of about two, comparable with (even lower than) the initial S/N in the original stack (Fig.~\ref{res:Sut}). Such an a posteriori selection cannot be used to improve the S/N of the final stacked signal. 

\begin{figure}[t]
	\centering
	\includegraphics[width=0.40\textwidth]{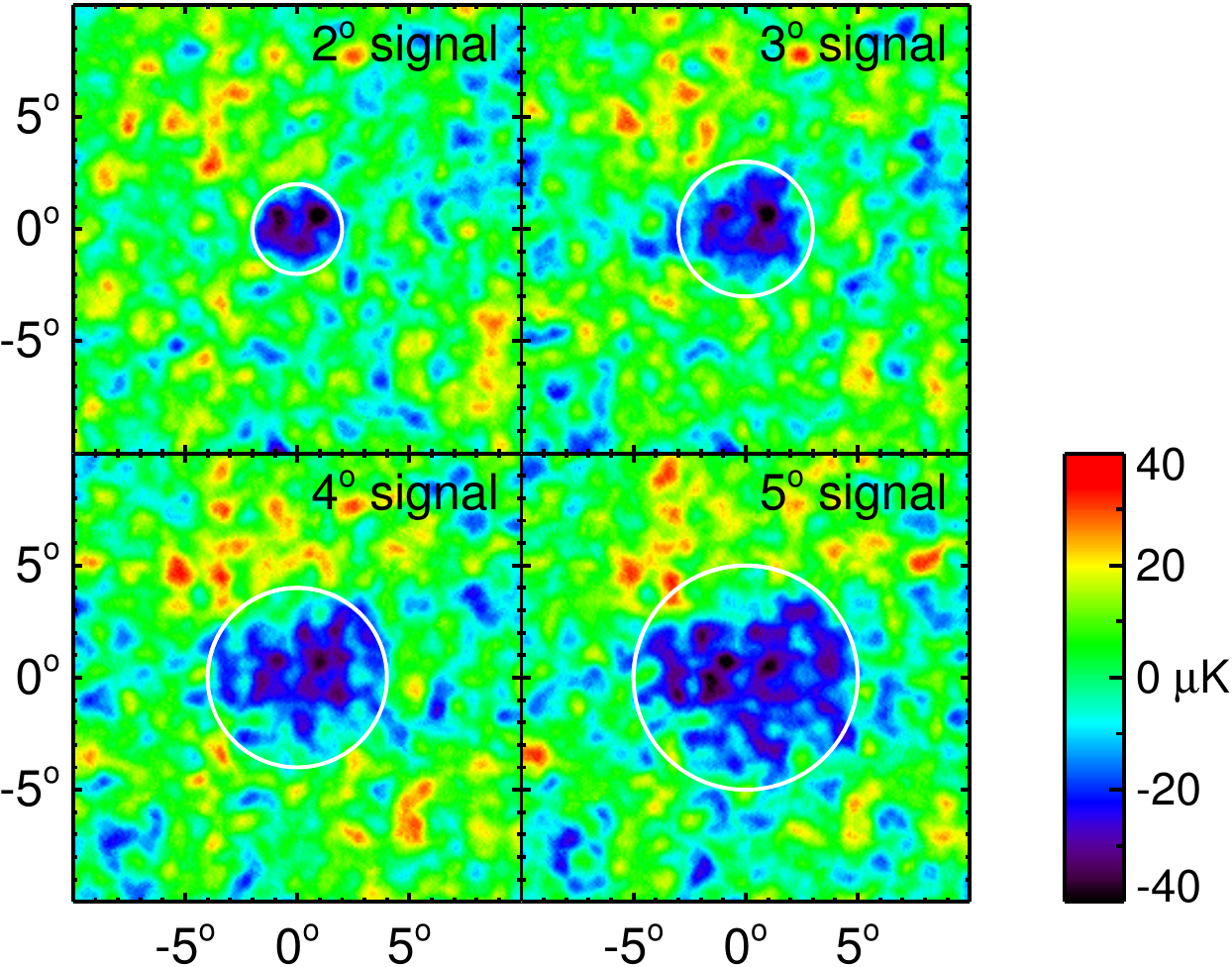}
	\caption{\small Illustration of selection effects. From a single set of around 200 random positions, we are able to construct a false ``iSW-like'' signal on any desired scale (here for angular radii of 2, 3, 4 and $5^\circ$) by selecting and stacking each time the appropriate half of the same set.}
	\label{res:selec_im}
\end{figure}

Actually, taking a closer look at these selection effects, we notice that with only one set of a few hundred random positions, one can obtain a strong -- but completely artificial -- signal on almost any desired scale ( Fig.~\ref{res:selec_im}) by selecting the appropriate half of it. This further illustrates, and warns us about the risk of a posteriori selections, and does put  any apparently significant signal we obtained into perspective. Our results from the new catalogues can be considered safe, since the only form of selection comes from the division of the Sutter et al.\ voids into redshift subsamples that has already been performed a priori by the authors. The Gr08 results are probably also safe, although their 50 voids come originally from a ten times larger sample (see Sect.~\ref{sec:2-2}) and were selected according to a density-contrast criterium. This selection was also made by the authors prior to looking at CMB data. It is not clear, however, whether that selection might have helped increasing the S/N artificially. Of further interest is the observation made by the Gr08 that either increasing or decreasing by a few tens the amount of voids in their selected sample does make the significance drop. This may either indicate a possible selection effect -- thus casting doubt on the iSW nature of the signal -- or be an expected effect due to the addition of noise-dominated voids (when increasing the number of voids) or the deletion of contributing voids (when decreasing it). 

\subsection{Alignment \& overlap effects}
\label{sec:5-2}

When interpreting our results, another important issue arises due to the number and location of the voids studied. In principle, each one of them leaves an imprint in the CMB temperature. These hundreds of voids are confined in the area covered by the SDSS. Since the angular size of these objects often exceeds several degrees, they are bound to fill this area and overlap. As a consequence, the stacking of a chosen set of voids in fact contains  contributions from many others. Some structures may also be very close to each other on the sky, further complicating the interpretation of signals. 

\noindent We devise two ways of quantifying these issues:
\begin{itemize}
	\item First, for each individual sample, we compute both the total area covered by all the voids and the area where at least two voids overlap; we then compute the ratio of these quantities, which represents the ``self-contamination" of each sample.
	\item Then, for every possible pair of samples, we compute the fraction of the area of the first sample shared by the second sample area -- a measure of the contamination between samples.
\end{itemize}
\noindent The results are compiled in Table~\ref{tab:align}. We can see that each sample shows moderate to high contamination with other samples and overlaps itself quite strongly. The results are lower for the Gr08 sample due to the small number of voids. Still, the first row of Table~\ref{tab:align} shows that many voids from other catalogues will contribute to the stacking of the Gr08 voids, making it more difficult to interpret its relatively high significance results. This is true for every catalogue of voids identified in the same area. The level of overlap is very high, and it would be difficult to determine which are the objects that produce the actual iSW signal. Moreover, the proximity of the voids may artificially amplify some detections due to the possibility of repeated stacking on the location of a significant signal/void.

Both aspects of this study show that contribution from other samples is indeed frequent in the stacking of these voids. Supposedly, we can expect the contribution to cancel out for large numbers of stacked patches, although this argument may be weakened by the fact that voids represent a much larger fraction of the volume of the Universe than high overdensities. In any case, the probability of measuring a false signal (due to a fortuitous event) may be heightened by the overlapping of so many voids. An accurate simulation of the expected signature from such a collection of overlapping structures will help quantify these contaminations and their effects on a possible iSW effect detection.

\begin{table}[t]
	\caption{\small Summary of the overlapping study for the void samples.}  
	\centering
	\begin{tabular}{CCCCCCCCCC}
		\hline
		\hline		
		\ & \ & \multicolumn{8}{|C}{\textbf{Fraction of the surface occupied by... }} \\
		\ & \ & \multicolumn{1}{|C}{\textbf{Gr08}} & \textbf{Pan} & \textbf{Sut-1} & \textbf{Sut-2} & \textbf{Sut-3} & \textbf{Sut-4} & \textbf{Sut-5} & \textbf{Sut-6} \\
		\hline
		\Tab{ \rotatebox{90}{\hspace{-2.5cm}\textbf{...in common with ...}}} &
		\Tab{\textbf{Gr08} \\ \textbf{Pan} \\ \textbf{Sut-1} \\ \textbf{Sut-2} \\ \textbf{Sut-3} \\ \textbf{Sut-4} \\ \textbf{Sut-5} \\ \textbf{Sut-6} } &
		\Tab{ \multicolumn{1}{|C}{\ }     \\   \multicolumn{1}{|C}{100\%}  \\   \multicolumn{1}{|C}{88.3\%}  \\   \multicolumn{1}{|C}{86.0\%}  \\   \multicolumn{1}{|C}{82.1\%}  \\   \multicolumn{1}{|C}{70.6\%}  \\   \multicolumn{1}{|C}{85.4\%}  \\   \multicolumn{1}{|C}{36.3\%}} &
        \Tab{14.3\%  \\    \     \\  47.8\%  \\  45.1\%  \\  41.9\%  \\  35.8\%  \\  45.1\%  \\  24.1\%} &
        \Tab{26.3\%  \\  99.3\%  \\     \     \\  85.4\%  \\  79.9\%  \\  69.3\%  \\  84.6\%  \\  46.0\%} &
        \Tab{27.3\%  \\  100\%  \\  91.3\%  \\    \      \\  88.1\%  \\  75.0\%  \\  92.4\%  \\  51.2\%} &
        \Tab{28.1\%  \\  100\%  \\  91.9\%  \\  94.8\%  \\    \      \\  77.0\%  \\  93.7\%  \\  51.8\%} &
        \Tab{28.2\%  \\  100\%  \\  93.2\%  \\  94.5\%  \\  90.1\%  \\     \     \\  94.2\%  \\  51.4\%} &
        \Tab{27.1\%  \\  100\%  \\  90.3\%  \\  92.3\%  \\  87.0\%  \\  74.7\%  \\    \      \\  50.7\%} &
        \Tab{21.6\%  \\  100\%  \\  92.1\%  \\  95.9\%  \\  90.1\%  \\  76.5\%  \\  95.1\%  \\    \    } \\
		\hline
		\ & \ & \multicolumn{8}{|C}{\textbf{Fraction of overlapping zones}} \\
		\ & \ & \multicolumn{1}{|C}{13.5\%} & 81.8\% & 82.2\% & 81.5\% & 67.9\% & 49.9\% & 86.4\% & 25.3\% \\
		\hline
	\end{tabular}
	\label{tab:align}
	\tablefoot{\textit{First part (top):} each column indicates for each samples the fraction of its surface that is contaminated by every other sample. \textit{Second part (bottom line):} shows for each sample the fraction of its covered surface where 2 or more voids overlap.}
\end{table}

\subsection{Spurious detections?}
\label{sec:5-3}

We would also like to stress that the superstructure data sets should be considered with caution. While writing this paper, several updates of the Sutter et al.\ catalogue were released. The changes were stated as having no impact on the conclusions of the associated paper \citep{Sutter2012}. A detailed examination showed that modifications ranged from minor to more consequent. Numerous additional small voids were detected thanks to an improved void finding algorithm. Quite a few voids were removed from one version to the next, and many others have seen modifications in their redshifts, sizes, and positions on sky, up to the point where one of the subsample was almost completely different (\emph{lrgbright}). These updates did, however, have a significant impact on the stacking of the voids, especially through inclusion of the new small voids. For instance, with the July version, our procedure yielded a $\sim 2.5\,\sigma$ detection of a negative signal at about $2.4^\circ$ in the \emph{bright2} subsample prior rescaling, while in the latest version of the same subsample the photometry shows a positive excess with almost $2\,\sigma$ significance. Conversely, rescaling the voids improved the detection of a negative photometric decrement only with the latest version of the catalogue. While the present status of the catalogue should be considered as robust (P. M. Sutter, private communication), we feel that the fact mentioned above makes it rather difficult to interpret without ambiguity the signal obtained.

11026

\section{Conclusion}
\label{sec:CCL}

In this paper, we revisited the stacking of voids in CMB maps as a potential probe of dark energy through the expected iSW effect from these structures. Previous work by \citet[][Gr08]{Gran2008} measured a $3.7\,\sigma$ signal in WMAP 5 ILC map using a catalogue of 50 selected supervoids extracted from the SDSS (DR6). We devised a complete protocol for a stacking procedure from a careful choice of maps to a rigorous estimation of the significance. We first applied it to the catalogue of voids of Gr08 and did not find any significant difference, if not a little weaker signal and associated S/N (by $0.4\,\sigma$). We then extended the analysis to two new void catalogues by \citet{Pan2012} and \citet{Sutter2012}. The first did not yield any significant result, most likely owing to the limited range of redshift and radii of the sample. 

The second new catalogue, however, hinted at more significant signals (although not nearly as strong as the Gr08 results) with a trend for the preferred scale in the signal, which seemed to point to half the mean size of the voids used in the stacking. This was not found with the Gr08 voids, for which the highest significance scale was close to the mean void size. We note, however, in favour of our results from the Sutter et al. catalogue that because of the irregular geometry of the stacked voids, we expect the preferred characteristic scale to in fact be noticeably smaller than the void themselves. Additionally, we showed that the rescaled signal from the Gr08 voids prefers a scale that is larger than the size of these objects, a feature that is rather hard to explain physically, unless these central voids are surrounded by unrealistically high overdensities. These results are in close agreement with the analysis performed by \citet{Planck_ISW} using the recently released CMB data. 

The rescaling of the CMB patches (according to the void sizes) prior to stacking proved to be a mandatory step toward obtaining a significant signal in the photometry profiles, especially in light of our results with the incremental stacking of the largest voids of the Sutter et al. catalogue. However, an unexpected signal with moderately high significance was found on small scales in these same profiles (and to a lesser extent with the Gr08 voids), which could be attributed to random CMB fluctuations, the void irregular shapes, and uncertainties in the position of the voids actual centres. A similarly high signal was also measured on high scales; however, we showed that a large portion of it originated in large scale fluctuations ($2<\ell<20$) in the CMB map, unrelated to the smaller scales that we are investigating.

Along with these results, we also addressed the risks of possible selection effects that could easily lead to an overestimation of the significance. We also stressed that the surface density of the voids within the SDSS area make them overlap significantly, making it even more difficult to formulate clear expectations about, and interpretations of, the measured signals. Finally, it is known that voids are actually difficult to identify with certainty and that one must proceed with caution when analysing such void samples. Another instance of this is that while being identified in the same SDSS DR7 set, the void subsamples \textit{dim1} and \textit{dim2} of Sutter et al. lie within the same redshift range as the voids identified by Pan et al., but they cover quite a different range in size and are distributed differently in redshift. We argue, therefore, that, combined with the unavoidable overlap of voids along a line of sight mentioned above, any claim of a detection of an iSW-like signal by the stacking of voids and/or claim of an oddity with respect to $\Lambda$CDM would be premature.

\begin{acknowledgements}
We are happy to acknowledge discussions with C. Hern\'andez-Monteagudo and J. M. Diego at an early stage of this work. We also thank N. Aghanim and F. Lacasa for stimulating discussions. Clarifying exchanges with the authors of \citet{Cai2013} and with P. M. Sutter are also acknowledged. Some of the results in this paper have been derived using the HEALPix \citep{Gorski2005} package. The work of S. Ili\'c is funded by the Doctoral Programme `AAP 2010 contrats doctoraux Paris-Sud 11'.
\end{acknowledgements}

\bibliographystyle{aa}
\bibliography{references_isw,isw_stack_voids,Planck_bib}

\appendix

\end{document}